\documentclass[letterpaper,twocolumn,10pt]{article}
\usepackage{usenix2019_v3}

\usepackage{tikz}
\usepackage{amsmath}

\usepackage{filecontents}

\usepackage{tabularx}
\usepackage{booktabs}
\usepackage{ragged2e}
\usepackage{amsmath}
\usepackage{amssymb}
\usepackage{enumitem}
\usepackage{times}
\usepackage{hyperref}
\usepackage{rotating}
\usepackage{pdflscape}
\usepackage{breakurl}
\usepackage[style=default]{mdframed}
\usepackage{listings}
\usepackage{hyperref}
\usepackage{xurl}

\begin{document}

\makeatletter
\twocolumn[
\begin{@twocolumnfalse}

\begin{center}

\vspace*{-0.4cm}

{\fontsize{17}{25}\selectfont\bfseries
GARAGE: Characterizing the Automation Boundary in\\
LLM-based Attack Graph Generation
\par}

\vspace{1.3cm}

{\fontsize{13}{16}\selectfont
Daekwon Pi$^{1}$,
Sangho Lee$^{2}$,
Young Hun Lee$^{3}$,
and Huy Kang Kim$^{1}$
\par}

\vspace{0.45cm}

{\fontsize{10.5}{13}\selectfont

$^{1}$ School of Cybersecurity,
Korea University,
Seoul, Republic of Korea
\\[-0.5mm]
{\ttfamily\bfseries
daekp@korea.ac.kr,
cenda@korea.ac.kr
}

\vspace{0.15cm}

$^{2}$ Samsung Electronics,
Suwon, Republic of Korea
\\[-0.5mm]
{\ttfamily\bfseries
s35.lee@samsung.com
}

$^{3}$ Agency for Defense Development (ADD),
Daejeon, Republic of Korea
\\[-0.5mm]
{\ttfamily\bfseries
yh\_lee@add.re.kr
}
\par}

\end{center}

\vspace{0.9cm}

\begin{center}
\begin{minipage}{0.88\textwidth}

\fontsize{10.5}{13.5}\selectfont
\justifying

\noindent

\textbf{Abstract.} While modern vehicle security depends on effective Cyber Threat Intelligence (CTI) synthesis, current automated tools struggle with unstructured data and automotive-specific architectural nuances. To bridge this gap, we introduce GARAGE, a RAG-powered framework that converts fragmented CTI into an actionable, domain-specific knowledge base for automated attack graph generation. GARAGE synthesizes a dataset of 12,786 CVEs and 140 incident reports into a STIX 2.1 and Auto-ISAC ATM-compliant knowledge base. By formalizing tactical-pattern-level scenarios through granular kill chain analysis, GARAGE achieves threat generation capabilities. Our 320 Leave-One-Out experiments reveal that the framework can accurately transfer security knowledge to entirely unseen vehicle architectures. Furthermore, we position GARAGE as a scalable TARA support tool within human-in-the-loop workflows, offering a comprehensive cost-performance analysis to guide its deployment across various LLM tiers.

\vspace{0.45cm}

\enspace$\cdot$\enspace
Large Language Models
\enspace$\cdot$\enspace
Retrieval-Augmented Generation
\enspace$\cdot$\enspace
Vehicle Security
\enspace$\cdot$\enspace
Cybersecurity

\end{minipage}
\end{center}

\vspace{1.0cm}

\end{@twocolumnfalse}
]
\makeatother


\section{Introduction}
While Large Language Models (LLMs) offer significant potential for automated security analysis, their application in the automotive industry faces distinct challenges. Much of the knowledge required for cybersecurity analysis remains private, fragmented, and difficult to share across organizations. 

One promising application of LLMs to overcome this limitation is generating vehicle-level attack graphs. Traditional Threat Analysis and Risk Assessment (TARA) is often scoped to individual Electronic Control Units (ECUs). In practice, however, safety-critical attacks are rarely confined to a single ECU. Instead, they exploit chained vulnerabilities across multiple components and architectural layers, including in-vehicle networks (\textit{e.g.}, CAN/Ethernet), operating systems, and hardware interfaces. Constructing and validating such end-to-end attack paths requires cross-domain expertise spanning networking, OS internals, and hardware, which is difficult for human analysts to acquire and systematically apply. LLMs can bridge this gap by aggregating distributed technical knowledge to support multi-layer security reasoning.

Generating attack graphs requires modeling relationships between assets, vulnerabilities, and attack actions (\textit{e.g.}, penetration, exploitation). Graph-based approaches provide explicit structure but often fail to capture the semantic context embedded in unstructured reports. Vector-based retrieval captures semantic similarity but often fails to preserve causal, step-by-step chaining. In this work, we adopt a Hybrid Retrieval-Augmented Generation (Hybrid RAG~\cite{sarmah2024hybridrag}) approach that jointly leverages graph-based structural knowledge and vector-based semantic retrieval for attack graph construction.

This challenge is further exacerbated by the manual nature of current TARA practice~\cite{wang2024proactive}, which prolongs assessment duration and hinders timely incorporation of rapidly evolving, unstructured Cyber Threat Intelligence (CTI). To address this, LLMs can act as automated analysts, directly generating attack graphs and inferring vulnerability chaining to support actionable, vehicle-level security assessment~\cite{pham2021survey}.

In this paper, we present GARAGE (\textbf{G}enerative \textbf{A}I with \textbf{R}AG-based \textbf{A}ttack \textbf{G}raph \textbf{E}ngine), an end-to-end framework that transforms publicly available automotive CTI into a domain-specific Knowledge Base (KB) and automatically generates vehicle-level attack graphs via RAG-based inference. Our main contributions are:

\begin{itemize}
    \item We construct an automotive cybersecurity KB from 12,786 CVEs and 140 security event documents, aligned with STIX~2.1 and the Auto-ISAC Automotive Threat Matrix (ATM). An LLM-based pipeline extracts security-relevant entities, relations, and events to populate both a Knowledge Graph (KG) and a vector store. We publicly release this KB to support reproducibility and future research.\footnote{\url{https://anonymous.4open.science/r/GARAGE-dataset-4265/}}
    \item We design a dual-metric evaluation framework (Practical Feasibility and Knowledge Reconstruction) that separately assesses attack path viability and security knowledge depth, enabling per-dimension evaluation of LLM capabilities.
    \item Through 320 Leave-One-Out (LOO) experiments across four real-world attacks and eight LLMs, we empirically establish an \textit{automation boundary} based on \textit{knowledge granularity}. Specifically, LLMs can generate high-level tactical threat scenarios, whereas low-level implementation details require human expert intervention.
    \item We provide quantitative deployment guidance through cost-performance trade-off analysis across proprietary and open-weight model tiers, demonstrating the utility of GARAGE as a practical TARA support tool within a Human-in-the-loop workflow.
\end{itemize}

\section{Methodology}

\begin{figure*}[tbp]
    \centering
    \includegraphics[width=0.9\textwidth]{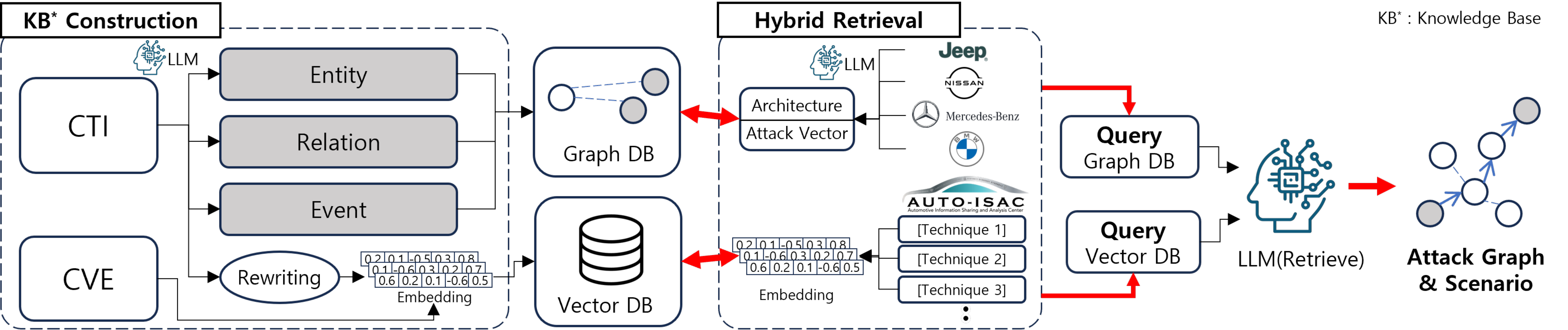}
    \caption{Proposed GARAGE System Architecture and Data Flow. The system integrates graph-based structural retrieval with vector-based semantic retrieval to improve attack graph generation quality.}
    \label{fig:full_width_architecture}
\end{figure*}

\subsection{Overview of the Pipeline}
    GARAGE is a Retrieval-Augmented Generation (RAG)-based system that automatically generates vehicle-level attack graphs. Inputs include unstructured CTI texts, CVE descriptions, ATM technique definitions, and (optionally) target vehicle architecture specifications. The output is an attack graph in structured JSON format that captures step-by-step vulnerability chaining, supported by evidence retrieved from both the vector store and KG.

    The pipeline combines two complementary databases. An LLM extracts security-relevant entities, relations, and events from CTI texts and stores them in a KG and a vector store. By integrating semantic similarity from the vector store with explicit structural connectivity across ECUs, components, and vulnerabilities stored in the KG, GARAGE captures both semantic context and structural chaining.

    During inference (Figure~\ref{fig:full_width_architecture}), GARAGE (i) formulates ATM-guided queries and performs vector retrieval to collect relevant CTI evidence, (ii) traverses the KG for high-confidence structural evidence around target assets/ECUs, and (iii) uses an LLM to synthesize both evidence sources into attack paths. To mitigate hallucinations, the LLM outputs are constrained to remain consistent with retrieved entities/relations and are supported by the underlying textual evidence.
    
\subsection{Dataset}
    The CTI dataset was sourced from the \textit{AutoSec-Timeline} repository (as of December 17, 2025)~\cite{autosec_timeline}, comprising \textbf{12,786 CVEs} and \textbf{140 security events}.

\subsubsection{Vulnerability Data}
    To filter CVEs relevant to automotive software, we collected software names and versions from the open-source license notices of three OEMs (Hyundai MOBIS~\cite{hd}, Volkswagen~\cite{vw_oss}, Kia~\cite{kia_oss}), selected for the public availability of their detailed software disclosures. Using this data, we applied regex-based filtering to identify relevant vulnerabilities. Selected vulnerability descriptions were stored in the vector store, and extracted entities were ingested into the KG (see Section~\ref{subsec:kb_construction}).
    
\subsubsection{Security Event Data}
    Raw textual data were manually curated from various sources, including technical blogs, reports, and presentations. To maintain format consistency, PDF and PPT documents were converted to text using the \texttt{pymupdf4llm}\ library, while YouTube content was processed via transcript extraction. For image-based materials, we provided manually authored natural-language descriptions. A total of 164 cases were initially gathered, of which \textbf{24} were excluded due to source inaccessibility (\textit{e.g.}, 404 errors, inaccessible Tor hidden services) or a lack of technical detail. This refinement process yielded \textbf{140} documents, whose event descriptions were subsequently embedded and indexed in the vector store.

\subsection{Knowledge Base Construction}
\label{subsec:kb_construction}
    All KB construction stages (entity, relation, event extraction, and rewriting) are independently performed by the same four LLMs: \textbf{Gemini 3.0 Flash, GPT-5.2, Claude 4.5 Sonnet, and DeepSeek-V3.2}. The goals are (i) to compare model-specific \textit{knowledge} and \textit{reasoning} characteristics in security analysis tasks, and (ii) to evaluate how evidence grounding via RAG enhances inference across model tiers.

\subsubsection{Entity extraction}
    We design an entity extraction pipeline to identify cybersecurity elements from automotive CTI sources. As inputs, we use the filtered CVE \texttt{Description} field and CTI documents segmented with \texttt{chunk\_size=500} and \texttt{chunk\_overlap=150}. Each chunk is processed independently, and results from the same document are merged at the document level.

    Each model extracts entity mentions under 12 predefined types (Table~\ref{tab:entity_types}). Results are stored in structured JSON and provided as the normalized entity set ($E_{extracted}$) to constrain the candidate space for relation extraction.

    \begin{table}[tbp]
      \centering
      \caption{The Entity Types for Automotive CTI}
      \label{tab:entity_types}
      \small
      \begin{tabularx}{\columnwidth}{@{}
        >{\ttfamily\RaggedRight}l
        >{\RaggedRight}X
        @{}}
        \toprule
        \textrm{\textbf{Type}} & \textbf{Definition} \\
        \midrule
        component & HW/SW module within the vehicle \\
        target\_ecu & Explicitly targeted ECU in an attack \\
        vulnerability & Identified security defect (incl.\ CWE) \\
        network\_domain & Vehicle network segment or logical domain \\
        network\_protocol & Communication protocol used in attack \\
        attack\_vector & Method of attack delivery \\
        supplier & Component supplier / manufacturer \\
        researcher & Researcher / org.\ that reported a vuln. \\
        cve & CVE identifier \\
        cwe & CWE identifier \\
        asset & Safety-critical function / system affected \\
        tool & HW/SW tool used to perform an attack \\
        \bottomrule
      \end{tabularx}
    \end{table}

    Our domain-specific entity schema resolves the semantic ambiguity of generic STIX objects (\textit{e.g.}, \texttt{Device}) that fail to distinguish automotive-specific components. By explicitly separating entities like IVI and Gateway ECU, it enables precise, domain-specific relations such as \texttt{(target\_ecu)--[BELONGS\_TO]--(network\_domain)}.

    All LLM engines share a standardized prompt framework with three core directives: (1) \textbf{Verbatim Extraction} to prevent hallucinations, (2) \textbf{Negative Instruction} to prohibit generating unmentioned entities, and (3) \textbf{Structural Guidance} with clear definitions for consistent classification (full prompt templates in Appendix~\ref{app:prompts}).

\subsubsection{Entity normalization}
    To eliminate semantic redundancy, we apply an embedding-based normalization. Raw entities are integrated with CVE product names, descriptions, and CWE information, then embedded using OpenAI's \texttt{text-embedding-3-large} ($\phi: E_{raw} \rightarrow \mathbb{R}^d$).
    We perform Agglomerative Hierarchical Clustering with Average Linkage and a cosine distance threshold $\tau = 0.3$, empirically tuned to balance under-clustering (\textit{e.g.}, failing to merge `CAN' and `Controller Area Network') against over-clustering of distinct entities.
    The shortest string in each cluster is selected as the canonical representative ($canonical(e) = \arg\min_{e' \in cluster(e)} |e'|$), preferring standard domain acronyms. In Neo4j, we adopt a \textbf{Canonical Node} architecture: representative nodes (\textit{e.g.}, \texttt{CanonicalComponent}) are created and linked from individual entities via \textbf{\texttt{SAME\_AS}} relationships, consolidating duplicate entries into canonical representations.

\subsubsection{Relation extraction}

    To identify semantic associations between extracted entities, we extract Subject-Predicate-Object (SPO) triplets $T = (e_s, r, e_o)$ from the CTI corpus $D$. The same four LLMs independently perform relation extraction, each receiving the raw text alongside the normalized entity list ($E_{extracted}$).

   \begin{table}[tbp]
      \centering
      \caption{The Predefined Relation Types for Triplet Extraction}
      \label{tab:relation_types}
      \small
      \begin{tabularx}{\columnwidth}{@{}
        >{\ttfamily\RaggedRight}l
        >{\RaggedRight}X
        @{}}
        \toprule
        \textrm{\textbf{Relation Type}} & \textbf{Definition} \\
        \midrule
        HAS\_VULNERABILITY & Component/ECU contains a vulnerability \\
        IDENTIFIED\_BY & Vulnerability is associated with a CVE \\
        CONNECTED\_TO & Component/ECU connects to a network domain \\
        USES\_PROTOCOL & Component/ECU/tool uses a network protocol \\
        DELIVERED\_VIA & Vulnerability is triggered via an attack vector \\
        PROVIDED\_BY & Supplier provides a component/ECU \\
        REPORTED\_BY & Researcher/org.\ reported a vulnerability \\
        AFFECTS\_ASSET & Vulnerability affects a specific asset \\
        USED\_AGAINST & Tool is used to attack a target \\
        \bottomrule
      \end{tabularx}
    \end{table}

    All prompts enforce five core constraints: (1) \textbf{Entity Constraint} ($e_s, e_o \in E_{extracted}$); (2) \textbf{Relational Constraint}, limiting $r$ to 9 predefined types (Table~\ref{tab:relation_types}); (3) \textbf{Verbatim Consistency}; (4) \textbf{Negative Instruction} to suppress hallucinations; and (5) \textbf{Structural Enforcement} via strict JSON schema. The resulting KG, averaged across the four models, comprises approximately \textbf{8,824} unique entity nodes and \textbf{9,265} relation edges.

\subsubsection{Event extraction}

    To model dynamic causality and behavioral patterns, we extract security events from unstructured text using the same four LLMs. Each model identifies security incidents and transforms them into structured event nodes comprising \textit{actor}, \textit{action}, \textit{target}, and \textit{pre/post-conditions}.
    Each extracted event is then mapped to specific ATM techniques via a fine-tuned classification model (detailed in Section~\ref{subsec:technique_extraction}), which assigns the corresponding \texttt{technique\_id} (\textit{e.g.}, ATM-T0059) and links it through a \texttt{CLASSIFIED\_AS} relationship. This hierarchical structuring allows tracing how individual actions on specific components map to broader attack tactics.

\subsubsection{Rewriting}
    The Rewriting phase converts structured entities and relations back into natural language to improve retrieval quality. Each of the four LLMs independently synthesizes factual summaries from verified fact triplets, excluding the noise present in raw CTI texts. The resulting summaries are stored in the vector store, directing the RAG system to retrieve refined factual information rather than noisy unstructured data (Appendix~\ref{app:prompts}).

\subsubsection{Technique extraction}
\label{subsec:technique_extraction}
    To classify ATM techniques, we developed a fine-tuned model via Supervised Fine-Tuning (SFT) based on \texttt{GPT-4.1-2025-04-14}. Zero-shot prompting showed high costs and limited precision for granular technique classification, necessitating the SFT approach. The training dataset was constructed from the latest ATM data with virtual incident reports synthesized in five narrative styles (research reports, advisories, press releases, forum posts, and training materials; Appendix~\ref{app:prompts}).
    Three measures ensure training robustness: (1) label leakage prevention, which blacklists technique names and IDs from generated narratives; (2) data augmentation via character-level noise injection and random truncation; and (3) semantic consistency validation, where each augmented sample is re-classified by a separate LLM to filter inconsistent examples. Additionally, negative samples drawn from non-security contexts (\textit{e.g.}, vehicle maintenance logs) were included to suppress false positives.

\subsection{Attack graph generation using Hybrid RAG}

    \subsubsection{Context-Aware Vector Retrieval}
    To reflect specific vehicle specifications in the retrieval process, we employ a three-stage retrieval pipeline. (1)~\textbf{Dynamic Query Generation}: the system analyzes the Vehicle Specification to identify entities such as ECUs and protocols. (2)~\textbf{ATM-Guided Query Formulation}: queries are constructed by concatenating each ATM technique's title and description to align retrieval with operational contexts. (3)~\textbf{Cross-Encoder Reranking}: initial candidates are re-ranked using a Cross-Encoder (ms-marco-MiniLM-L-6-v2) that evaluates query-document pairs simultaneously to select the top-10 documents as final context.

    \subsubsection{Structured Knowledge Graph Traversal}
    To supplement unstructured text with explicit component connectivity, we employ the Neo4j-based KG. ECU IDs and protocol names from the vehicle spec are mapped as \textbf{Seed Nodes}, and traversal is strictly filtered to follow only security-critical relations (\texttt{HAS\_VULNERABILITY}, \texttt{DELIVERED\_VIA}, \texttt{EXPLOITS}) with a maximum depth of 2-hops. This constraint effectively captures core causal chains such as ``ECU $\rightarrow$ Component $\rightarrow$ Vulnerability,'' avoiding exhaustive graph traversal.
    For densely-connected nodes, raw SPO triples are replaced by an LLM-generated security summary that compiles the entity's associated vulnerabilities and attack vectors, following the community summarization principle of GraphRAG~\cite{edge2024local}. This reduces prompt token consumption while preserving essential security semantics.

    \subsubsection{Hybrid Context Integration and Attack Graph Generation}
    Similar attack cases retrieved from the vector store and component-vulnerability relationships traversed from the KG are integrated into a unified prompt. When the two sources conflict, a conservative approach is adopted: vulnerabilities recorded in the KG are assumed to exist by default. Vector store evidence (\textit{e.g.}, patch notes) overrides this assumption only when it explicitly references the target vehicle's make, model, and year, in which case the feasibility of the corresponding attack path is reduced or excluded. The LLM then uses this hybrid context together with the target vehicle's network topology to infer multi-stage attack paths, outputting the result as a structured attack graph.

\section{Experiment}

We designed experiments around three research questions to evaluate the effectiveness, reliability, and practical utility of GARAGE.

\begin{itemize}
\item \textbf{RQ1: Inference Capability on Unknown Threats.}
Evaluating whether the framework can infer potential attack paths solely from vehicle architectural specifications when direct knowledge of the target attack is absent.

\item \textbf{RQ2: Reliability as an Expert Assistant.}
Determining whether the framework maintains consistent results across repeated trials, ensuring the stability required for an automated expert assistant.

\item \textbf{RQ3: Cost-Performance Efficiency.}
Investigating whether open-weight models provide a viable, cost-effective alternative to proprietary models while maintaining sufficient analytical depth.
\end{itemize}


\subsection{Evaluation Benchmarks}

We selected four well-documented automotive cyber incidents ($C_1$--$C_4$) representing diverse architectures and attack vectors:

\begin{itemize}
    \item \textbf{$C_1$: Jeep Cherokee\cite{jeep_hacking}} — Remote exploitation via cellular network targeting Uconnect and gateway (V850) to control CAN bus. Validates attack paths in legacy architectures.

    \item \textbf{$C_2$: Mercedes-Benz W177\cite{mercedes_hacking}} — Exploitation of head unit--backend communication and WebKit flaws. Evaluates connected car analysis.

    \item \textbf{$C_3$: Nissan Leaf\cite{nissan_hacking}} — Bluetooth stack buffer overflow via the Hands-Free Profile (HFP) for RCE on IVI. Tests short-range wireless vulnerability detection.

    \item \textbf{$C_4$: BMW i3\cite{bmw_hacking}} — Multi-vector attacks (GSM, Bluetooth, USB) compromising HU and TCU, with gateway bypass via UDS diagnostic messages.
\end{itemize}

\noindent \textbf{Ground Truth \& Controls.}
We manually reconstructed attack scenarios from detailed technical white papers, referencing ATM incident examples for $C_1$ and $C_4$~\cite{autoisac_atm}. Ground truth was established by mapping attack paths to ATM Tactics and Techniques.

Two strict controls were enforced: (1) a LOO strategy excluding target incident data from the KB, and (2) \textbf{Entity Anonymization} to prevent parametric memory recall and force architectural reasoning. We omitted vehicle make and model names from queries, and strictly filtered out retrieved results from both the vector store and KG that contained target-specific keywords (\textit{e.g.}, \texttt{jeep}, \texttt{uconnect} for $C_1$; \texttt{bmw}, \texttt{connecteddrive} for $C_4$). General technical terms (\textit{e.g.}, SPI, CAN, Bluetooth) are intentionally retained as domain-common knowledge shared across vehicles.


\subsection{Experimental Environment}

\subsubsection{Knowledge Base Configuration}
\label{sub:kb_config}
To evaluate end-to-end capability, we constructed independent KBs using four SOTA models (GPT-5.2, Gemini 3.0 Flash, Claude 4.5 Sonnet, DeepSeek-V3.2) as Extraction Models.

Since each model produces a unique knowledge representation, we created 16 hybrid KB instances ($4$ Extraction Models $\times$ $4$ scenarios), each with LOO applied to both stores:

\begin{itemize}
    \item \textbf{Graph DB (Neo4j):} Graph topology varies across instances because each LLM extracts distinct triples.
    \item \textbf{Vector DB (FAISS):} Unlike standard RAG, our vector store is built via natural language rewriting of extracted triples, making embeddings model-dependent and requiring 16 separate indices.
\end{itemize}

The KGs contain an average of approximately 8,824 nodes and 9,265 edges. Vectorization uses OpenAI text-embedding-3-small (dim=1536).

\subsubsection{Target LLMs}
\label{sub:target_llms}

We selected eight LLMs including both proprietary and open-weight categories. For the experiment, models are grouped by their \textit{functional role} rather than license type:

\begin{itemize}
    \item \textbf{Extraction \& Inference (4 models):} GPT-5.2, Gemini 3.0 Flash, Claude 4.5 Sonnet, DeepSeek-V3.2, accessed via official APIs. These serve dual roles as both Extraction Models (Section~\ref{sub:kb_config}) and inference targets evaluated on their own KB instances. Note that DeepSeek-V3.2 is an open-weight model but is included in this group due to its competitive extraction capability and API accessibility.

    \item \textbf{Inference Only (4 models):} Qwen3-coder, Llama 4 Maverick, GPT-OSS-120b, GPT-OSS-20b, accessed via OpenRouter API\cite{openrouter} for the inference phase only.
\end{itemize}

\noindent \textbf{Evaluation Mapping.}
Extraction \& Inference models query their own KB instances, while Inference Only models are uniformly mapped to the GPT-5.2-generated KBs as the standard retrieval pool, selected as the most capable extraction model available at the time of experimentation.

\noindent In the Results section, models are regrouped by \textit{weight availability}---Proprietary (Claude, GPT, Gemini) vs.\ Open-Weight (DeepSeek, Qwen3, Llama, GPT-OSS)---to analyze performance differences attributable to model architecture and scale.

\noindent \textbf{Hyperparameters.}
Generation \texttt{temperature} is set to 1.0 for all models to encourage exploration of diverse multi-vector attack paths. The GARAGE framework constrains outputs with retrieved evidence, ensuring structural validity despite the elevated sampling temperature.


\subsection{Evaluation Methodology: LLM-as-a-Judge}
\label{sub:metrics}

A single metric cannot simultaneously capture attack-path viability and knowledge accuracy; the dual design enables separate analysis---for example, high KR but low PF indicates strong domain knowledge yet poor path consistency, an outcome commonly observed in open-weight models.

\subsubsection{Practical Feasibility (PF)}

PF evaluates attack path viability from a red-team perspective, enforcing the \textbf{cascade failure} principle: if step $k$ is invalidated, all subsequent steps ($i \ge k$) are scored as failed. PF comprises three components:

\noindent (1) \textbf{Chain Completeness (Max 60):} Ratio of contiguous valid steps before the first invalidation point $k$. Each step $s_i$ is scored as VALID~(1.0), PARTIAL~(0.5), or BROKEN~(0.0):
$$Score_{PF\text{-}Comp} = \frac{\sum_{i=1}^{N} s_i}{N} \times 60 \quad \text{where } s_i = 0 \text{ for } i \ge k$$
(2) \textbf{Attack Feasibility (Max 30):} Logical transition validity between adjacent steps, scored as FEASIBLE~(1.0), QUESTIONABLE~(0.5), or INFEASIBLE~(0.0):
$$Score_{PF\text{-}Feas} = \frac{\sum_{j=1}^{k-1} t_j}{N-1} \times 30$$
(3) \textbf{Critical Violations (up to $-$20):} Penalties for physical/logical impossibilities (\textit{e.g.}, trust boundary crossing $-10$, protocol mismatch $-5$). No double jeopardy with elements already scored~0.

\subsubsection{Knowledge Reconstruction (KR)}

Unlike PF, KR measures the depth of security knowledge independently of logical chain integrity---accurately identifying a subsequent step accrues points even if a preceding step is missed.

\noindent (1) \textbf{Entity \& Step Identification (Max 50):} A 5-level semantic matching scale ($m_i$): FULL~(1.0), SEMANTIC~(0.75), PARTIAL~(0.5), WEAK~(0.25), or NONE~(0.0):
$$Score_{KR\text{-}Iden} = \frac{\sum_{i=1}^{N} m_i}{N} \times 50$$
(2) \textbf{Structural Knowledge (Max 30):} Transition connectivity classified as CONNECTED~(1.0), IMPLIED~(0.5), or DISCONNECTED~(0.0):
$$Score_{KR\text{-}Struct} = \frac{\sum_{j=1}^{N-1} c_j}{N-1} \times 30$$
(3) \textbf{Knowledge Precision (Max 20):} Baseline 10 points; bonuses for accurate details (\textit{e.g.}, exact CVEs), deductions only for hallucinations (\textit{e.g.}, fabricated ECUs). Valid alternative attack paths are not penalized.

\vspace{0.5em}
\noindent \textbf{Evaluation Setup.}
We adopt the LLM-as-a-Judge approach, which recent studies have shown to achieve high agreement with human expert ratings when guided by structured scoring criteria~\cite{zheng2023judging}. To mitigate self-evaluation bias, we employ Claude 4.5 Sonnet (temperature $=$ 0.3) as an independent Judge. The fixed scoring categories and explicit criteria in our PF/KR evaluation scheme further constrain subjective variation. The evaluation covers 320 instances (8 models $\times$ 4 scenarios $\times$ 10 iterations), yielding 640 evaluation profiles (PF + KR per instance).

\section{Results and Analysis}

This section analyzes 320 experiments (8 models $\times$ 4 cases $\times$ 10 iterations) conducted under strict LOO control. Appendix Figure~\ref{fig:ag_comparison_bmw} illustrates a representative comparison between a manually curated ground-truth attack graph and an LLM-generated counterpart; the best-scoring generated graphs for all four cases are provided in Appendix Figures~\ref{fig:best_ag_jeep}--\ref{fig:best_ag_bmw}.

\subsection{Inference Capability on Unknown Threats (Answering RQ1)}
\label{sub:rq1}

The main hypothesis of GARAGE is that attack paths for unknown threats can be reconstructed by \emph{transferring} knowledge from other vehicle attack cases, even when direct attack knowledge for the target vehicle is absent from the KB.
We conducted 320 experiments across 8 models under the LOO condition.
Proprietary models (Claude~4.5~Sonnet, GPT~5.2, Gemini~3.0~Flash) achieved a mean PF of 59.0/100 (64.8/55.9/56.3; KR~64.1), while Open-Weight models (Qwen3-Coder, DeepSeek-V3.2, Llama-4, GPT-OSS) achieved a mean PF of 39.2/100 (KR~48.8).

\begin{figure*}[t]
  \centering
  \includegraphics[width=0.9\textwidth]{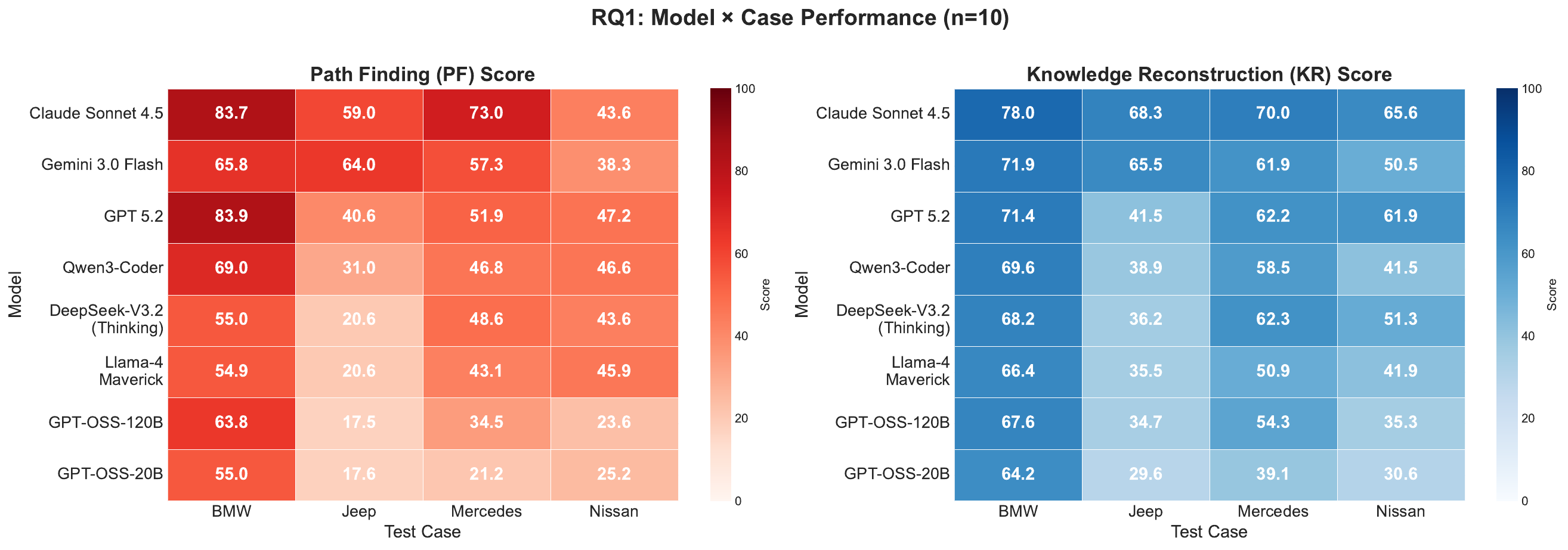}
  \caption{PF/KR Score Heatmap across 8 models and 4 attack cases. Proprietary models (top three rows) consistently occupy the higher score bands, the BMW column yields the highest scores across all models, and the Jeep column exhibits the largest variance.}
  \label{fig:heatmap}
\end{figure*}

\begin{figure}[t]
  \centering
  \includegraphics[width=\columnwidth]{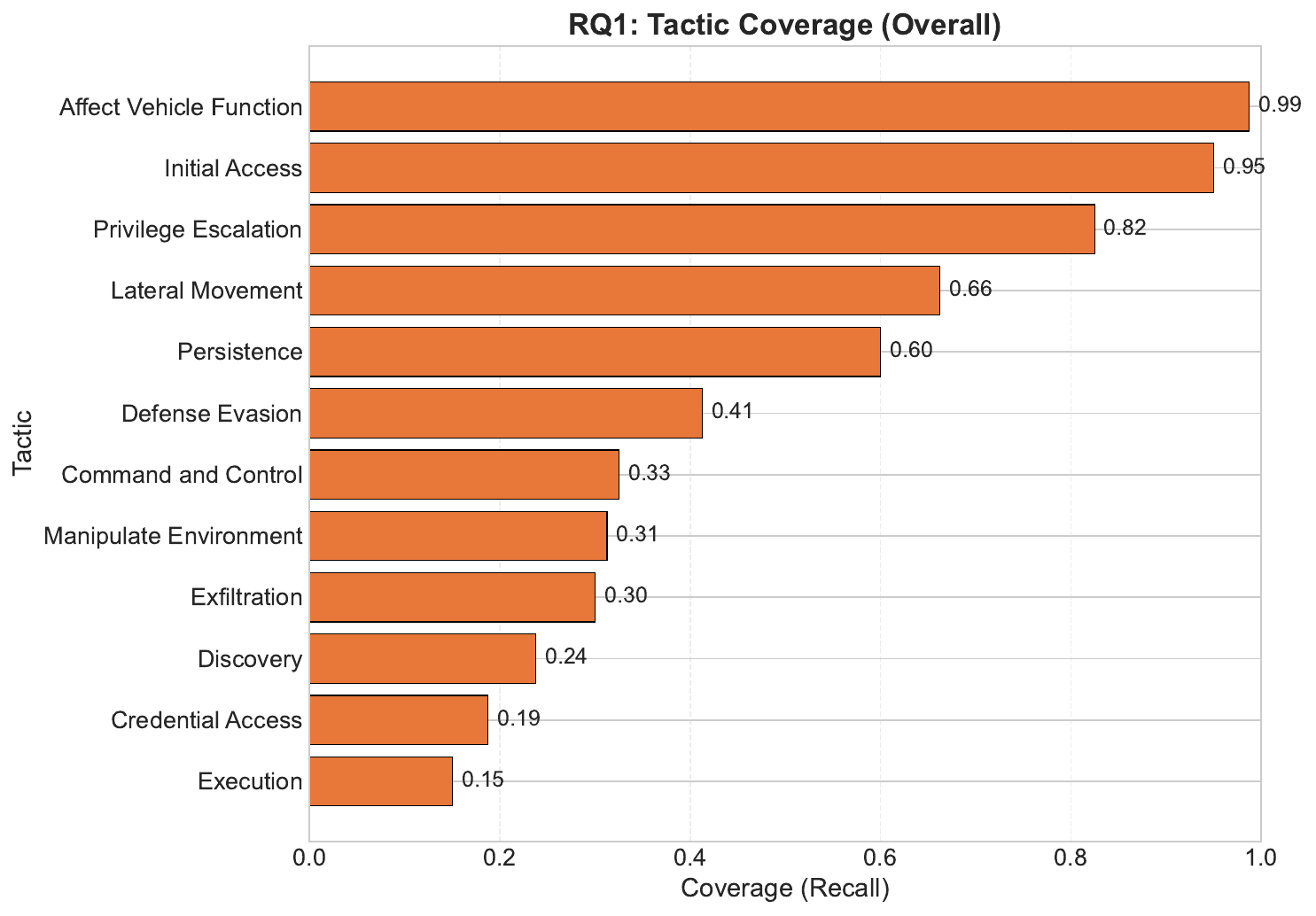}
  \caption{Auto-ISAC Kill Chain tactic coverage. Initial Access (95\%) and Affect Vehicle Function (99\%) show high coverage, while mid-chain steps (Execution~15\%, C2~33\%) drop sharply. This pattern is consistent across all 8 models.}
  \label{fig:killchain}
\end{figure}

    However, these aggregate scores alone cannot explain the inter-model and inter-case performance variations. As a RAG-based system, GARAGE's performance can be evaluated across two dimensions: (1)~the \textbf{Retrieval} stage, defined by the availability of transferable references in the KB; and (2)~the \textbf{Generation} stage, representing the LLM's accuracy in composing retrieved knowledge into executable attack paths.
    Section~\ref{sec:rq1-model-case} analyzes model- and case-level patterns from the heatmap (Figure~\ref{fig:heatmap}), and Section~\ref{sec:rq1-killchain} extends this analysis from a tactical perspective (Figure~\ref{fig:killchain}).

\subsubsection{Model and Case Analysis}
\label{sec:rq1-model-case}

We analyze the heatmap patterns based on four factors: \textit{knowledge granularity}, structural complexity, compositional reasoning ability, and RAG augmentation scope.

\paragraph{Observation 1: Knowledge granularity determines transfer success.}

    BMW i3 achieved the highest scores across all 8 models (55.0--83.9), whereas Jeep Cherokee yielded the lowest (17.5--64.0).
    To investigate this variance, we analyzed the \texttt{match\_status} labels assigned by the Judge's KR module.
    The \textbf{KB Match Rate} (the ratio of FULL or SEMANTIC verdicts) represents the overlap between the GT abstraction level and KB coverage. A higher rate indicates that the GT is defined at the tactical-pattern level, where existing KB threat patterns provide sufficient references. A lower rate indicates that the GT requires target-specific implementation details not covered by cross-vehicle knowledge.

    \begin{table}[t]
    \centering
    \caption{KB Match Rate and mean PF score by attack case. Higher KB Match Rate---reflecting GT definitions covered by existing KB threat patterns---correlates with higher PF.}
    \label{tab:kb-match-rate}
    \resizebox{\columnwidth}{!}{%
    \begin{tabular}{lccp{3.5cm}}
    \toprule
    \textbf{Case} & \textbf{KB Match} & \textbf{Mean PF} & \textbf{GT Examples} \\
                  & \textbf{Rate (\%)} &                 & \textbf{(key\_entities)} \\
    \midrule
    BMW     & \textbf{88.8} & 66.4 & USB Interface, \texttt{Central Gateway}, CAN Bus \\
    Benz    & \textbf{53.4} & 47.1 & \texttt{Code Execution}, \texttt{Privilege Escalation} \\
    Jeep    & \textbf{38.8} & 33.9 & \texttt{/fs/mmc0}, \texttt{cmcioc.bin}, \texttt{0xF0 0x02} \\
    Nissan  & \textbf{35.4} & 39.2 & \texttt{CVE-2017-7932}, \texttt{libevo\_stack.so}, \texttt{dnscat2} \\
    \bottomrule
    \end{tabular}%
    }
    \end{table}

    These results indicate two key findings.
    First, GARAGE is \textbf{most effective for threat modeling defined at the tactical-pattern level}.
    Second, in high KB Match Rate environments (BMW, 88.8\%), performance gaps between models narrow. For example, GPT-OSS-20B (PF~55.0) achieved performance comparable to proprietary models.
    Notably, despite similar KB Match Rates between Nissan (35.4\%) and Jeep (38.8\%), PF scores differed (39.2 vs.\ 33.9), indicating that structural complexity (Observation~2) and model reasoning ability (Observations~3--4) also significantly affect performance.

\paragraph{Observation 2: Model capability divergence under reference sparsity.}

    The Jeep Cherokee case exhibited a substantial performance gap between Proprietary and Open-Weight models, with the SPI-based V850 firmware flashing step as the primary bottleneck.

    \begin{table}[t]
    \centering
    \caption{Jeep Cherokee: Per-model PF scores and step-level VALID rates. The SPI/firmware step serves as the key discriminator between Proprietary and Open-Weight models.}
    \label{tab:jeep-bifurcation}
    \resizebox{\columnwidth}{!}{%
    \begin{tabular}{lcccc}
    \toprule
    \textbf{Model} & \textbf{Jeep PF} & \textbf{VALID Rate} & \textbf{SPI/FW} & \textbf{Persist.} \\
    \midrule
    Gemini   & \textbf{64.0} & \textbf{55\%} & \textbf{4/10} & 0/10 \\
    Claude   & \textbf{59.0} & \textbf{46\%} & \textbf{5/10} & 0/10 \\
    GPT      & 40.6          & 5\%           & 0/10          & 0/10 \\
    Qwen3    & 31.0          & 8\%           & 0/10          & 0/10 \\
    DeepSeek & 20.6          & 8\%           & 0/10          & 0/10 \\
    Llama    & 20.6          & 0\%           & 0/10          & 0/10 \\
    OSS-20B  & 17.6          & 0\%           & 0/10          & 0/10 \\
    OSS-120B & 17.5          & 0\%           & 0/10          & 0/10 \\
    \bottomrule
    \end{tabular}%
    }
    \end{table}

    This step involves hardware-dependent lateral movement not observed in other KB cases. Claude and Gemini successfully inferred paths by combining ATM techniques (\textit{e.g.}, ATM-T0068 Firmware Manipulation) with SPI protocol references. For instance, Claude generated ``Reflash V850 firmware via compromised SPI,'' adapting generic firmware techniques to the target architecture (VALID 4--5/10). All other models failed at this reasoning task.
    The Persistence step, which relies on target-specific filesystem structures (\texttt{/fs/mmc0}), recorded 0/10 VALID across all models, confirming that knowledge granularity constrains the upper bound of model performance.

\paragraph{Observation 3: Architecture recognition and cascade failure.}

    The three Proprietary models recorded similar scores on BMW (Claude~83.7, GPT~83.9, Gemini~65.8) but diverged sharply on Jeep (Gemini~64.0, Claude~59.0, GPT~40.6).
    GPT dropped 43.3 points on Jeep despite achieving the overall highest score on BMW, because Jeep's GT requires system-specific protocols (D-Bus, SPI) that its generic-pattern strategy could not handle.

    \begin{table}[t]
    \centering
    \caption{Jeep Cherokee step-level evaluation for Proprietary models. V\,=\,VALID, P\,=\,PARTIAL, B\,=\,BROKEN (counts out of 10 iterations).}
    \label{tab:jeep-step-level}
    \resizebox{\columnwidth}{!}{%
    \begin{tabular}{clccc}
    \toprule
    \textbf{Step} & \textbf{Phase} & \textbf{Gemini} & \textbf{Claude} & \textbf{GPT} \\
    \midrule
    1 & Reconnaissance        & \textbf{10}/0/0 & \textbf{9}/0/1 & 1/2/\textbf{7} \\
    2 & Initial Access        & \textbf{10}/0/0 & \textbf{9}/0/1 & 1/0/\textbf{9} \\
    3 & HU Compromise (RCE)   & 7/3/0           & 4/4/2          & 0/5/\textbf{5} \\
    4 & Persistence           & 0/\textbf{7}/3  & 0/5/5          & 0/1/\textbf{9} \\
    5 & Lateral Movement (FW) & \textbf{4}/1/5  & \textbf{5}/0/5 & 0/1/\textbf{9} \\
    6 & Command Translation   & \textbf{5}/0/5  & \textbf{5}/0/5 & 0/1/\textbf{9} \\
    7 & Steering Control      & 3/2/5           & 0/5/5          & 1/2/7 \\
    8 & Braking/Engine        & 5/2/3           & 5/1/4          & 1/3/6 \\
    \bottomrule
    \end{tabular}%
    }
    \end{table}

    GPT's performance drop originated in the reference-rich early steps (Steps~1--2): the model applied generic HTTP/TCP 80/443 patterns in 9/10 iterations instead of the D-Bus protocol (Port~6667) specified in the GT. Due to PF's cascade failure property, this initial error nullified all subsequent scores.
    In contrast, Gemini and Claude achieved 9--10/10 VALID on Steps~1--2 and maintained 4--5/10 VALID on sparse intermediate steps by formulating alternative mechanisms.
    This outcome indicates that \textbf{accurate target architecture recognition is a prerequisite for overall path quality}, and that architecture adaptation ability varies significantly even among Proprietary models (cross-case SD: GPT~16.7 vs.\ Gemini~10.9).

\paragraph{Observation 4: Structural reasoning---parallel vs.\ sequential paths.}

    On the Benz case, Claude~4.5~Sonnet scored 73.0, a gap of 15.7+ points over the second-tier group (Gemini~57.3, GPT~51.8).
    The Benz scenario involves \textbf{two independent attack chains} (T-Box and Head Unit) converging at the CAN bus.

    \begin{table}[t]
    \centering
    \caption{Benz case: VALID rates for critical steps across two parallel attack paths (Head Unit and T-Box).}
    \label{tab:mbux-parallel}
    \resizebox{\columnwidth}{!}{%
    \begin{tabular}{lccccc}
    \toprule
    \textbf{Model} & \textbf{VALID} & \textbf{HU IA} & \textbf{HU CAN} & \textbf{T-Box IA} & \textbf{T-Box CAN} \\
                   & \textbf{Rate}  &                & \textbf{-HMI}   &                   & \textbf{-D} \\
    \midrule
    Claude & \textbf{64\%} & 10/10 & \textbf{10/10} & 9/10  & \textbf{8/10} \\
    Gemini & 48\%          & 9/10  & 6/10           & 9/10  & 4/10 \\
    GPT    & 41\%          & 6/10  & \textbf{1/10}  & \textbf{10/10} & 5/10 \\
    \bottomrule
    \end{tabular}%
    }
    \end{table}

    Claude maintained high VALID rates across both paths. GPT, however, generated the T-Box path successfully (10/10) but failed to construct the Head Unit path (1/10).
    Interestingly, this structural difficulty caused a \textbf{performance reversal} among lower-tier models (Figure~\ref{fig:heatmap}): Llama scored Nissan~45.9 vs.\ Benz~43.1, and GPT-OSS-20B scored Nissan~25.2 vs.\ Benz~21.2.
    Because Nissan employs a \textbf{sequential single-path} attack chain (via Bluetooth), lower-tier models processed this linear structure more effectively than Benz's parallel construct, despite Nissan's lower KB Match Rate (35.4\% vs.\ 53.4\%).
    This finding highlights that model performance depends not only on capability tier but also on \textbf{the structural complexity of the attack paths}.

\paragraph{Observation 5: RAG augmentation effect and PF--KR gap.}

\begin{table}[t]
\centering
\caption{PF--KR gap across all 8 models. The gap serves as a proxy for compositional reasoning---the ability to convert retrieved knowledge into executable attack paths.}
\label{tab:pf-kr-gap}
\resizebox{\columnwidth}{!}{%
\begin{tabular}{llccr}
\toprule
\textbf{Tier} & \textbf{Model} & \textbf{PF} & \textbf{KR} & \textbf{Gap} \\
\midrule
Proprietary  & Claude 4.5 Sonnet  & 64.8 & 70.5 & +5.7 \\
Proprietary  & Gemini 3.0 Flash   & 56.3 & 62.4 & +6.1 \\
Proprietary  & GPT 5.2            & 55.9 & 59.3 & +3.4 \\
\midrule
Open-Weight & Qwen3-coder        & 48.3 & 52.1 & +3.8 \\
Open-Weight & DeepSeek-V3.2      & 42.0 & 54.5 & +12.5 \\
Open-Weight & Llama-4            & 41.1 & 48.7 & +7.6 \\
Open-Weight & GPT-OSS-120B       & 34.9 & 48.0 & +13.1 \\
Open-Weight & GPT-OSS-20B        & 29.8 & 40.9 & +11.1 \\
\bottomrule
\end{tabular}%
}
\end{table}

The PF--KR gap ranges from +3.4 to +13.1 across models (Table~\ref{tab:pf-kr-gap}), indicating a continuous range of compositional reasoning ability rather than a strict tier-based separation. This gap arises because KR credits individual step recognition independently of chain continuity, while PF's cascade failure penalizes broken causal chains. Most Open-Weight models exhibit large gaps (+7.6 to +13.1), indicating sufficient entity-level knowledge to recognize attack components but insufficient compositional reasoning to chain them into valid sequences. Notably, Qwen3-coder (+3.8) achieves Proprietary-level compositional efficiency despite being an Open-Weight model. As KB Match Rate decreases, these inter-model reasoning differences re-emerge as the dominant performance factor.

\paragraph{Summary.}
Based on the five observations, GARAGE's LOO performance primarily depends on four interacting factors: (1)~\textit{knowledge granularity}---the intersection of GT abstraction and KB coverage determines transfer success (Obs.~1); (2)~\textit{structural complexity}---parallel vs.\ sequential path topology affects reasoning difficulty (Obs.~4); (3)~\textit{model-specific compositional reasoning}---capability differences in architecture recognition and hardware-dependent inference become apparent (Obs.~2,~3); and (4)~\textit{RAG augmentation scope}---the degree to which retrieved knowledge is converted into executable paths (Obs.~5). The logical completion rate across all 320 experiments was 55.6\% (178/320), correlating strongly with knowledge granularity.

\subsubsection{Kill Chain Coverage Analysis}
\label{sec:rq1-killchain}

Coverage is computed as the mean rate at which each tactic was scored VALID or PARTIAL across all 320 experiments. The resulting distribution reveals distinct differences in transferability: Initial Access (95\%), Privilege Escalation (82.5\%), and Affect Vehicle Function (98.8\%) exhibit high coverage because their underlying entry vectors, escalation methods, and CAN-bus-based control principles share common patterns across vehicle architectures.
In contrast, Execution (15\%), C2 (32.5\%), and Persistence (60\%) recorded significantly lower coverage due to their reliance on target-specific implementations.
For example, the Execution tactic in $C_3$ (Nissan) necessitates triggering a buffer overflow in \texttt{libevo\_stack.so} via the HFP Bluetooth profile---a library exclusive to Nissan's IVI platform that generic references cannot anticipate. Similarly, C2 in $C_1$ (Jeep) relies on D-Bus IPC communicating via \texttt{NavTrailService} on port~6667, which is unique to Chrysler's Uconnect system. Identifying such specific elements requires reverse-engineering artifacts that are inherently absent from cross-vehicle KBs.
This pattern demonstrates the limitations of automated knowledge transfer. For steps dependent on target-specific firmware or software internals, relying solely on knowledge retrieval is insufficient, and reverse engineering or dynamic analysis via Human-in-the-loop validation remains necessary.

\begin{mdframed}[backgroundcolor=gray!10, linewidth=0.5pt]
\textbf{RQ1 Conclusion:}
GARAGE demonstrates that under the LOO condition, it can reconstruct valid attack paths (mean PF~59.0/100 for Proprietary models) by transferring knowledge from other vehicle attack cases and CTI.
Performance depends on four key factors: knowledge granularity, structural complexity, compositional reasoning, and RAG augmentation scope.
GARAGE is most effective for \textbf{tactical-pattern-level threat modeling}; implementation details present fundamental limitations that require Human-in-the-loop approaches.
\end{mdframed}


\subsection{Consistency Analysis (Answering RQ2)}
\label{sub:rq2}

\begin{figure*}[tbp]
    \centering
    \includegraphics[width=0.9\textwidth]{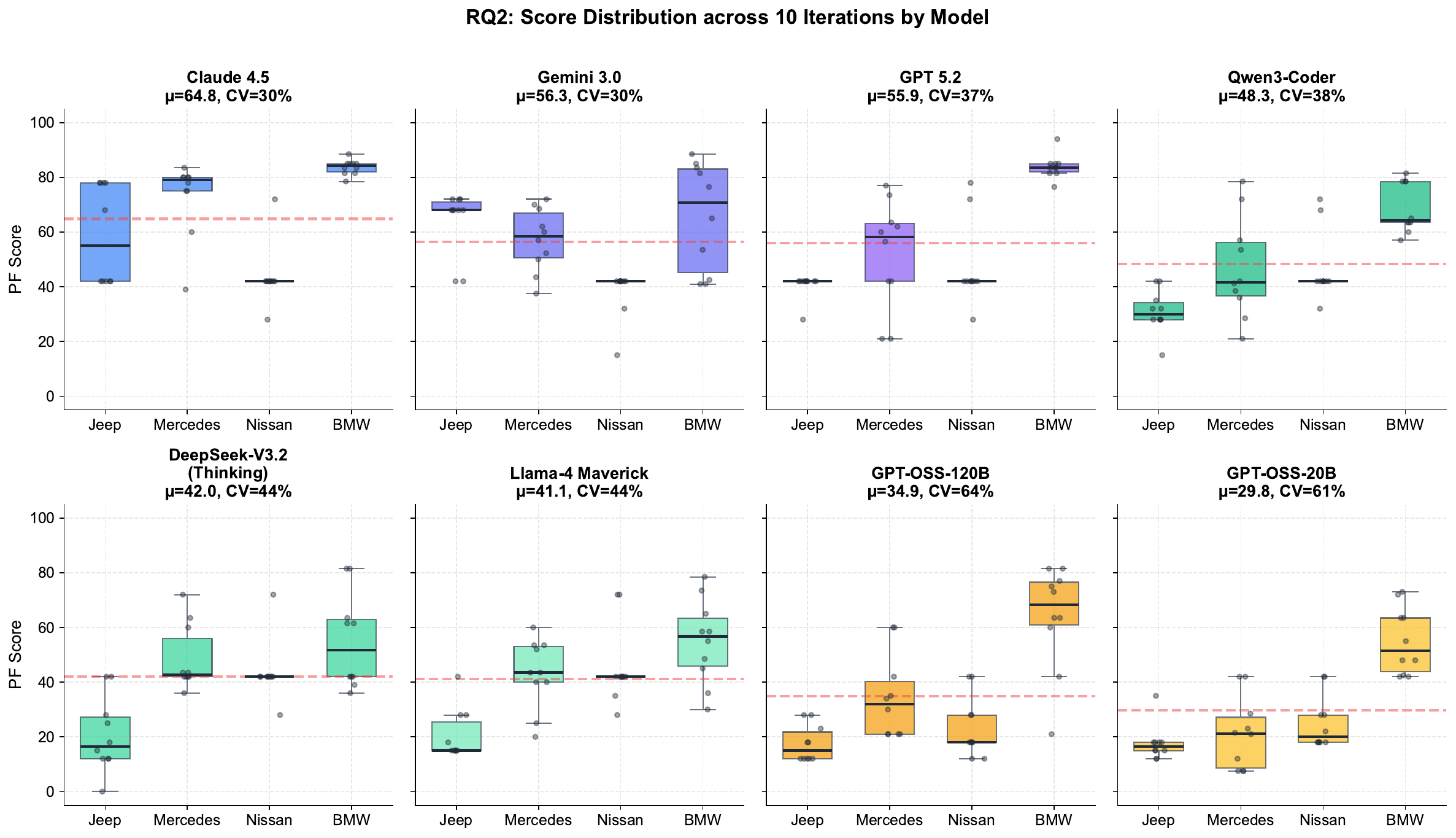}
    \caption{Distribution of PF scores across 10 independent trials for each model and case. Proprietary models (Claude, Gemini, GPT) exhibit narrower interquartile ranges, indicating higher consistency, while Open-Weight models show wider variance.}
    \label{fig:consistency}
\end{figure*}

To verify whether the performance differences observed in RQ1 are reproducible, all experiments were repeated 10 times and analyzed using the Coefficient of Variation (CV) and 95\% Confidence Interval (CI).
Figure~\ref{fig:consistency} visualizes the per-case PF distributions for each model: Proprietary models exhibit compact interquartile ranges concentrated in the upper score bands, while several Open-Weight models display wider spreads with data points frequently clustering near zero---illustrating the impact of cascade failures analyzed below.

\begin{table}[t]
\centering
\caption{PF score consistency metrics across 8 models (10 iterations $\times$ 4 cases = 40 experiments per model).}
\label{tab:rq2-consistency}
\resizebox{\columnwidth}{!}{%
\begin{tabular}{lrrrr}
\toprule
\textbf{Model} & \textbf{Mean (PF)} & \textbf{Std} & \textbf{CV (\%)} & \textbf{95\% CI} \\
\midrule
Claude 4.5 Sonnet & 64.8 & 19.5 & \textbf{30.1} & $\pm$6.0 \\
Gemini 3.0 Flash  & 56.3 & 17.1 & 30.3           & $\pm$5.3 \\
GPT 5.2           & 55.9 & 20.9 & 37.5           & $\pm$6.5 \\
Qwen3-coder       & 48.3 & 18.3 & 37.9           & $\pm$5.7 \\
DeepSeek-V3.2       & 42.0 & 18.6 & 44.3           & $\pm$5.8 \\
Llama-4           & 41.1 & 18.0 & 43.8           & $\pm$5.6 \\
GPT-OSS-120B      & 34.9 & 22.4 & 64.3           & $\pm$7.0 \\
GPT-OSS-20B       & 29.8 & 18.2 & 61.2           & $\pm$5.6 \\
\bottomrule
\end{tabular}%
}
\end{table}

Proprietary models exhibited a mean CV of 32.6\%, substantially lower than Open-Weight models (50.3\%).
The overall mean 95\% CI was $\pm$5.9 points, narrower than inter-model differences reported in RQ1 (\textit{e.g.}, Claude~64.8 vs.\ GPT~55.9, $\Delta\approx$9), confirming that the performance hierarchy is statistically meaningful and reproducible.
The elevated overall CV (43.7\%) reflects PF's bimodal score distribution from cascade failures rather than true instability. Specifically, PF's cascade failure property---where a single broken early step nullifies all subsequent scores---produces two distinct score clusters: high scores when early steps succeed (allowing the full chain to be evaluated) and near-zero scores when they fail (invalidating the entire path). This bimodality artificially increases variance-based metrics like CV without indicating actual output inconsistency; accordingly, the 95\% CI serves as the more appropriate reliability indicator.

\begin{mdframed}[backgroundcolor=gray!10, linewidth=0.5pt]
\textbf{RQ2 Conclusion:}
GARAGE exhibits reproducibility with a mean 95\% CI of $\pm$5.9 points. Proprietary models (CV$\approx$33\%) demonstrate higher consistency than Open-Weight models (CV$\approx$50\%).
The performance hierarchy from RQ1 is a stable, reproducible phenomenon.
\end{mdframed}


\subsection{Cost-Performance Trade-off (Answering RQ3)}
\label{sec:rq3}

To assess practical deployability, we analyzed the trade-off between per-analysis cost and performance across 8 models.
Costs were calculated from actual token counts measured from experimental prompts and outputs.

\begin{table}[t]
\centering
\caption{Per-analysis cost and performance metrics. Cost is computed from measured token counts and API pricing (as of Feb.\ 2026). Efficiency is defined as PF score per milli-dollar.}
\label{tab:rq3-cost}
\resizebox{\columnwidth}{!}{%
\begin{tabular}{llrrrrr}
\toprule
\textbf{Model} & \textbf{Tier} & \textbf{In Tok} & \textbf{Out Tok} & \textbf{Cost} & \textbf{PF} & \textbf{Eff} \\
               &               &                 &                  & \textbf{(\$/run)} &          & \textbf{(PF/m\$)} \\
\midrule
Claude 4.5 Sonnet     & Proprietary  & 5{,}942 & 4{,}990 & \$0.093 & 64.8 & 700 \\
GPT 5.2        & Proprietary  & 5{,}946 & 4{,}005 & \$0.066 & 55.9 & 841 \\
Gemini 3.0 Flash     & Proprietary  & 5{,}837 & 1{,}867 & \$0.009 & 56.3 & \textbf{6{,}614} \\
Qwen3-coder    & Open-Weight & 5{,}704 & 2{,}548 & \$0.004 & 48.3 & 12{,}712 \\
DeepSeek-V3.2    & Open-Weight & 3{,}756 & 3{,}195 & \$0.001 & 42.0 & 29{,}007 \\
Llama-4        & Open-Weight & 5{,}786 & 1{,}783 & \$0.002 & 41.1 & 21{,}217 \\
GPT-OSS-120B   & Open-Weight & 5{,}836 & 3{,}110 & \$0.001 & 34.9 & 42{,}578 \\
GPT-OSS-20B    & Open-Weight & 5{,}745 & 3{,}145 & \$0.001 & 29.8 & 48{,}560 \\
\bottomrule
\end{tabular}%
}
\end{table}

\begin{figure}[tbp]
    \centering
    \includegraphics[width=\columnwidth]{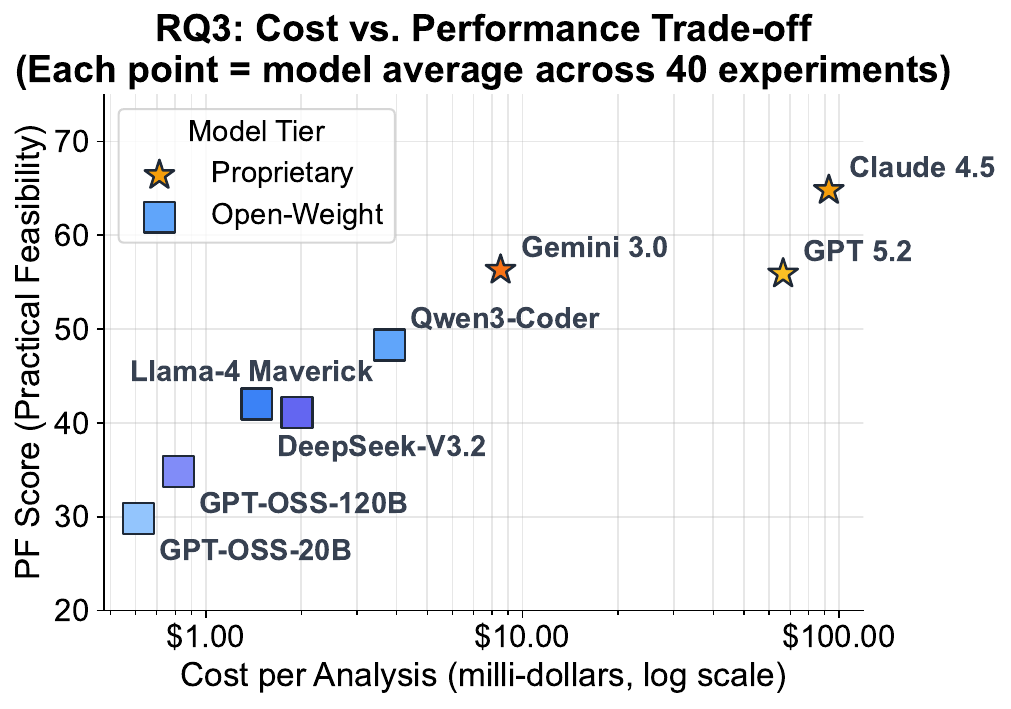}
    \caption{Cost vs. Performance trade-off across eight LLMs (log scale). Gemini 3.0 Flash achieves the optimal cost-performance balance among proprietary models, while Qwen3-coder represents the best value among open-weight alternatives.}
    \label{fig:cost_perf}
\end{figure}

    The cost-performance distribution segments into two tiers (Figure~\ref{fig:cost_perf}).
    \textbf{Within the Proprietary tier}, cost and performance do not scale linearly: Gemini~3.0~Flash achieved 87\% of Claude's performance at 1/10 the cost (\$0.009 vs.\ \$0.093), yielding 9.4$\times$ higher cost efficiency.
\textbf{In the Open-Weight tier}, Qwen3-coder (PF~48.3, \$0.004) achieved the highest performance at 1/17--1/24 the cost of Claude/GPT, while DeepSeek-V3.2 (PF~42.0, \$0.001) maintained meaningful performance at minimal cost. Smaller models (GPT-OSS) have the lowest API costs ($\le$\$0.001) but PF scores of 30--35; all Open-Weight models can be deployed on-premise, though GPU infrastructure costs would apply.

\begin{mdframed}[backgroundcolor=gray!10, linewidth=0.5pt]
\textbf{RQ3 Conclusion:}
GARAGE's cost-performance trade-off is clearly stratified by model tier.
Gemini~3.0~Flash provides the optimal cost-efficiency among Proprietary models (1/10 the cost of Claude at 87\% performance), while Qwen3-coder achieves the best cost-effectiveness among Open-Weight models.
\end{mdframed}

\section{Discussion and Limitation}

\subsection{Discussion}

\textbf{Key Findings.}
Our 320 LOO experiments across four real-world automotive attacks and eight LLMs reveal three key findings.
First, \textbf{knowledge-transfer-based inference on unknown threats is practically feasible}: even with all direct attack knowledge removed, proprietary models reconstructed valid attack paths with a mean PF of~59.0/100. This transfer occurs through two complementary pathways---explicit retrieval from the domain-specific KB and implicit recall from the LLM's parametric knowledge. While our ablation study shows that both pathways achieve comparable PF scores, only the KB-grounded pathway provides formal evidence traceability, a prerequisite for regulatory compliance (see \textit{Role of Retrieval Modality}).
Second, \textbf{performance is structurally constrained by knowledge granularity}: the strong correlation between KB Match Rate and PF (BMW 88.8\%$\rightarrow$66.4 vs.\ Jeep 38.8\%$\rightarrow$33.9) indicates that the abstraction level at which threats are defined sets an upper bound on automated inference. This constraint persists regardless of model capability. Tactical-pattern-level threats are highly suitable for automated reasoning, whereas implementation-detail-level threats establish a structural boundary for knowledge transfer.
Third, \textbf{the impact of model selection is highly context-dependent}: the best-worst performance gap (Claude 64.8 vs.\ GPT-OSS-20B 29.8) far exceeds the 95\% CI ($\pm$5.9), yet this variance is amplified in scenarios characterized by low knowledge granularity and high structural complexity. This suggests that the choice of model must be tailored to the specific characteristics of the threat type.

\textbf{Practical Implications.}
As an automated tool supporting TARA compliance under UN~R155 and ISO/SAE~21434, GARAGE can systematically generate initial threat scenario candidates. The high Kill Chain coverage in Initial Access (95.0\%) and Affect Vehicle Function (98.8\%), as reported in Section~\ref{sec:rq1-killchain}, indicates that the system can cover entry points and impact scenarios that analysts might otherwise overlook. Conversely, the lower coverage in Execution (15.0\%) and C2 (32.5\%) confirms that GARAGE is most effective as an \textit{expert assistant} within a Human-in-the-loop workflow, automating initial threat exploration while leaving technical verification and prioritization to domain experts.

In the context of the ISO/SAE~21434 TARA workflow, this boundary of automation maps directly to specific process stages: GARAGE facilitates \textit{threat scenario enumeration} and \textit{initial attack path generation}, corresponding to high-coverage tactics. However, \textit{attack feasibility validation} requires implementation-level knowledge of the target system (involving low-coverage tactics such as Execution and C2) and therefore remains a manual expert task.

For deployment, our RQ3 results provide quantitative criteria for model selection: Claude 4.5 Sonnet (PF~64.8) is suitable for accuracy-critical regulatory compliance; Gemini 3.0 Flash achieves 87\% of Claude's performance (PF~56.3 vs.~64.8) at a fraction of the cost, making it practical for large-scale analyses; and Qwen3-Coder is an alternative for air-gapped on-premise environments, though total cost of ownership (GPU infrastructure) should be considered beyond API pricing.

\textbf{Role of Retrieval Modality.}
An ablation study comparing Graph-only, Vector-only, and No-RAG conditions against the full Hybrid configuration showed negligible PF variations ($\Delta$PF $<$ 1.2 across all conditions). This indicates that \textbf{knowledge granularity---the quality and coverage of the underlying KB---is the dominant factor} dictating inference performance, rather than the specific retrieval modality. The KB provides the foundational knowledge that all retrieval strategies draw upon; accordingly, the core contributions of this work are the domain-specific KB construction pipeline, the dual-metric evaluation framework, and the empirical findings on knowledge transfer boundaries.

Nevertheless, the domain-specific KB and RAG pipeline are indispensable for practical deployment. While the No-RAG condition attains comparable PF scores by relying on parametric memory, its outputs lack \textbf{formal evidence traceability}---the ability to trace each attack step back to a specific CVE, CTI report, or KG node. Under UN~R155 and ISO/SAE~21434, outputs without formal grounding cannot serve as auditable compliance evidence. By design, the RAG-based architecture grounds generated attack paths in retrieved source artifacts, thereby providing the traceability demanded by regulatory standards. Quantifying the grounding rate of each generated step remains future work.

\subsection{Limitation}

\textbf{L1: Limited GT Scale and Subjectivity.}
The evaluation relies on four public automotive attack cases that are among the most thoroughly documented in automotive cybersecurity. However, the small sample size limits statistical generalizability. Because GT construction inevitably involves researcher interpretation, three domain experts independently reconstructed each GT by referencing ATM official examples ($C_1$, $C_4$) or technical white papers ($C_2$, $C_3$), resolving any disagreements through consensus. We further mitigated subjective bias by evaluating structural coherence and practical feasibility rather than demanding exact matches, thereby accepting semantically equivalent alternative paths.

\textbf{L2: LLM-as-a-Judge Bias.}
Utilizing Claude 4.5 Sonnet as both a target model and an evaluator introduces potential model-family bias. To mitigate this risk, we implemented four distinct safeguards: (i)~complete session isolation between generation and evaluation phases; (ii)~dual independent metrics (PF/KR); (iii)~uniform evaluation criteria across all tested models; and (iv)~a structured output format requiring all models to produce attack graphs in an identical JSON schema, eliminating stylistic preferences as a confounding factor. Empirically, no systemic bias was observed; for instance, Gemini (64.0) outperformed Claude (59.0) in the Jeep case.
To further validate judge reliability, we conducted cross-judge evaluation on two cases ($C_1$, $C_4$; $N$=160) using GPT-5.2 and Gemini 3.0 Flash. As detailed in Table~\ref{tab:cross-judge}, pairwise Spearman correlations indicated strong agreement ($\rho$ = 0.74--0.84, $p < 0.001$), confirming that model rankings remain consistent regardless of the judge model employed.

\begin{table}[h]
\centering
\caption{Cross-judge agreement ($N$=160, two cases combined). All correlations are significant at $p < 0.001$.}
\label{tab:cross-judge}
\begin{tabular}{lcc}
\toprule
\textbf{Judge Pair} & \textbf{PF $\rho$} & \textbf{KR $\rho$} \\
\midrule
Claude vs GPT    & 0.76 & 0.77 \\
Claude vs Gemini & 0.75 & 0.74 \\
GPT vs Gemini    & 0.82 & 0.84 \\
\bottomrule
\end{tabular}
\end{table}

\textbf{L3: LLM Hallucination.}
Across all 320 experiments, 76.9\% of the generated attack graphs contained at least one hallucinated element (averaging 1.74 per experiment), with hallucination rates ranging from 52.5\% (GPT~5.2) to 95.0\% (Qwen3-Coder). Common hallucination types include fabricated CVE identifiers, non-existent protocol names, and generic network details not grounded in the input vehicle specification. The mean Knowledge Precision score of 9.81 out of 20 indicates that approximately half of the generated knowledge elements lack direct traceability to the retrieved context. This result supports the need for a Human-in-the-loop workflow: while GARAGE effectively generates candidate attack paths, expert review remains essential to filter hallucinated artifacts before operational use.

\textbf{L4: LOO vs.\ True Zero-Day.}
The LOO experimental design simulates unknown threats by systematically excluding the target case; however, true zero-day scenarios may involve entirely novel attack surfaces that inherently lack any transferable references within the KB. Because our LOO condition ensures the availability of at least three other authoritative vehicle attack cases, it represents a controlled evaluation of knowledge transfer capability rather than a strict worst-case assessment.

\textbf{L5: Vehicle Architecture Input Requirements.}
Our experiments reconstruct vehicle architecture specifications from publicly available technical reports and white papers, representing the maximum detail recoverable from open sources. In practice, OEMs rarely disclose complete E/E architecture details or SBOMs, limiting specification fidelity for external analysts. Furthermore, validating operational exploitability would require physical testbed access. GARAGE is therefore designed to generate \textit{plausible threat scenarios} for TARA prioritization rather than confirmed exploit chains.

\textbf{L6: Generalizability Across Architectures.}
Our evaluation is based on vehicle architectures from the 2014--2018 era. Modern software-defined vehicles incorporate stronger security controls (e.g., HSMs, SecOC, and Automotive Ethernet) and may exhibit different transferability characteristics. However, the framework can be extended to newer architectures by updating the KB and threat taxonomy.


\section{Related Work}

\textbf{Attack Graphs and Automotive Domain Application.} AGs are a key tool for modeling multi-step attack paths. Early logic-based frameworks like MulVAL\cite{ou2005mulval} inferred attack paths from formalized vulnerability and configuration inputs. However, these approaches rely on predefined static rules, creating a knowledge acquisition bottleneck that limits integration of unstructured threat intelligence. Research applying AGs to the automotive domain primarily focuses on automating the TARA process mandated by the ISO/SAE 21434 standard. Early studies such as Salfer \textit{et al}.\cite{salfer2024automotive} and GAPP\cite{saulaiman2025graph} used AGs to analyze architectural vulnerabilities in on-board networks. Subsequent tools like ThreatGet\cite{chlup2022threatget} supported ISO/SAE 21434 compliance by automating asset-driven attack tree analysis. Separately, dynamic AG generation algorithms based on ontology reasoning were proposed for Internet of Vehicles environments. Nevertheless, these approaches often focus on vehicle-level assessment, showing limitations in addressing the 'function-level' TARA required by ISO/SAE 21434. Furthermore, the unique protocols (\textit{e.g.}, CAN, V2X) and architectures of vehicles create a domain knowledge gap that general-purpose IT tools cannot capture. To address this, domain-specific knowledge bases such as CAKG\cite{yang2023cakg} (a vehicle-specific cybersecurity knowledge graph) and the Acti dataset (manually annotated automotive CTI reports)\cite{wang2025dataset} have been introduced. Despite these contributions, they do not provide an end-to-end solution for automatically linking this knowledge to AG generation for TARA. 

\textbf{CTI-based Knowledge Graph Construction and LLMs.}
Extracting structured knowledge from unstructured CTI has been extensively studied. Standardized ontologies like UCO\cite{syed2016uco} provided a schema for knowledge representation. Early NLP-based studies, such as Open-CyKG\cite{injy2021open} and AttacKG\cite{li2022attackg} employed specially trained neural models to extract entities and relations from CTI reports. However, these methods were inflexible, overfit to specific datasets and schemas, and incurred high retraining costs to adapt to new threats. Recent advancements in LLMs have provided flexible alternatives. LLMs use pre-trained knowledge to parse complex CTI text without task-specific fine-tuning. LLM-TIKG\cite{hu2024llm} and CTINexus\cite{cheng2025ctinexus} proposed flexible and adaptable CTI-to-KG pipelines using LLM in-context learning (ICL). Furthermore, CTIKG\cite{huang2024ctikg} introduced multi-agent and dual-memory architectures to mitigate LLM hallucinations. Together, these studies show that LLMs can reduce the knowledge acquisition bottleneck in CTI analysis. Benchmarks such as CTI-HAL\cite{della2025cti} further enable systematic performance validation.

\textbf{LLM-driven Attack Graph Automation.}
Recent work integrates LLMs directly into the AG generation logic, moving beyond CTI-to-KG preprocessing. AttacKG+\cite{zhang2025attackg+} enhances AG construction by utilizing LLMs to infer dependencies between attack nodes. This trend has led to multimodal approaches like MM-AttacKG\cite{zhang2025mm}, which analyzes both text and images (\textit{e.g.}, threat diagrams, screenshots) from CTI reports. Despite these advancements, a critical gap remains between these LLM-based methods and the specific TARA requirements of the automotive domain.

Specifically, the latest LLM-AG frameworks are domain-agnostic. They lack the built-in knowledge required to model the unique constraints of an automotive E/E architecture or the specific vulnerabilities of the CAN bus. Conversely, automotive domain-specific research, such as CAKG\cite{yang2023cakg} and the Acti dataset\cite{wang2025dataset}, focuses primarily on building foundational knowledge bases but does not extend to end-to-end attack graph generation.

In summary, no existing framework effectively combines the parsing flexibility of LLMs with automotive-specific CTI processing to automatically generate TARA-ready AGs. This paper presents GARAGE to address this gap. GARAGE is an end-to-end framework that uses LLM-based CTI extraction, a domain-specific automotive cybersecurity KB, and RAG-based inference to generate vehicle-level attack graphs. We evaluate this framework using a dual-metric approach via empirical analysis across eight LLMs.

\section{Conclusion}

This paper presented GARAGE, an end-to-end framework that transforms publicly available automotive CTI into a domain-specific KB and automatically generates vehicle-level attack graphs via RAG-based inference. The KB comprises 12,786 CVEs and 140 security events, structured following STIX~2.1 and ATM. We evaluated the generated AGs using a dual-metric framework (PF and KR).

Through 320 LOO experiments across four real-world attack cases and eight LLMs, we demonstrated cross-vehicle knowledge transfer to unseen architectures. Specifically, proprietary models reconstructed valid attack paths with a mean PF of 59.0, even without direct knowledge of the target attacks. Based on kill chain analysis and \textit{knowledge granularity}, we identified a clear automation boundary: automated generation is effective for tactical-pattern-level threat scenarios, whereas implementation-detail-level stages require human expert intervention. An ablation study confirmed that KB quality and coverage, rather than retrieval modality, is the dominant performance factor, underscoring the centrality of the KB construction pipeline. Our primary contributions are therefore the domain-specific KB construction pipeline, the evaluation methodology, and the empirical findings regarding the automation boundary.

These results show that GARAGE serves as a practical TARA support tool within a human-in-the-loop workflow. Additionally, our findings provide cost-performance guidance for deploying various LLM tiers. Future work will extend the evaluation to software-defined vehicle architectures and explore integration with real-time CTI feeds.

\bibliographystyle{plain}
\bibliography{reference/reference}

@inproceedings{sarmah2024hybridrag,
  title={Hybridrag: Integrating knowledge graphs and vector retrieval augmented generation for efficient information extraction},
  author={Sarmah, Bhaskarjit and Mehta, Dhagash and Hall, Benika and Rao, Rohan and Patel, Sunil and Pasquali, Stefano},
  booktitle={Proceedings of the 5th ACM International Conference on AI in Finance},
  pages={608--616},
  year={2024}
}

@article{zheng2023judging,
  title={Judging llm-as-a-judge with mt-bench and chatbot arena},
  author={Zheng, Lianmin and Chiang, Wei-Lin and Sheng, Ying and Zhuang, Siyuan and Wu, Zhanghao and Zhuang, Yonghao and Lin, Zi and Li, Zhuohan and Li, Dacheng and Xing, Eric and others},
  journal={Advances in neural information processing systems},
  volume={36},
  pages={46595--46623},
  year={2023}
}

@article{saulaiman2025graph,
  title={Graph-based automation of threat analysis and risk assessment for automotive security},
  author={Saulaiman, Mera Nizam-Edden and Kozlovszky, Miklos and Csilling, Akos},
  journal={Information},
  volume={16},
  number={6},
  pages={449},
  year={2025},
  publisher={MDPI}
}

@misc{jeep_hacking,
  title = {Jeep Hacking},
  howpublished = {\url{https://www.ioactive.com/wp-content/uploads/pdfs/IOActive_Remote_Car_Hacking.pdf}},
  author={IOActive},
  note = {Accessed: 25 Feb 2026}
}

@misc{bmw_hacking,
  title = {Experimental Security Assessment of BMW Cars},
  howpublished = {\url{https://keenlab.tencent.com/en/whitepapers/Experimental_Security_Assessment_of_BMW_Cars_by_KeenLab.pdf}},
  author={KeenLab},
  note = {Accessed: 25 Feb 2026}
}

@misc{nissan_hacking,
  title = {Remote Exploitation of Nissan Leaf},
  howpublished = {\url{https://i.blackhat.com/Asia-25/Asia-25-Evdokimov-Remote-Exploitation-of-Nissan-Leaf.pdf}},
  author = {{Black Hat}},
  note = {Accessed: 25 Feb 2026}
}

@misc{mercedes_hacking,
  title = {Mercedes Benz Security Research Report},
  howpublished = {\url{https://keenlab.tencent.com/en/whitepapers/Mercedes_Benz_Security_Research_Report_Final.pdf}},
  author={KeenLab},
  note = {Accessed: 25 Feb 2026}
}

@misc{hd,
  title = {Hyundai MOBIS Open Sources},
  howpublished = {\url{https://www.mobis.com/kr/tech/rnd.do}},
  author={Hyundai MOBIS},
  note = {Accessed: 25 Feb 2026}
}

@misc{vw_oss,
  title = {Volkswagen third-party licence notes},
  howpublished = {\url{https://www.volkswagen-newsroom.com/en/third-party-licence-notes-38}},
  author = {Volkswagen},
  note = {Accessed: 25 Feb 2026}
}

@misc{kia_oss,
  title = {KIA open source software notice},
  howpublished = {\url{http://webmanual.kia.com/STD_GEN5_WIDE/AVNT/EU/Italian/opensourcesoftwarenotice.html}},
  author = {KIA},
  note = {Accessed: 25 Feb 2026}
}

@misc{autosec_timeline,
  title = {AutoSec-Timeline},
  howpublished = {\url{https://autosec-timeline.delikely.eu.org/}},
  author = {AutoSec},
  note = {Accessed: 25 Feb 2026}
}

@inproceedings{ou2005mulval,
  title={MulVAL: A Logic-based Network Security Analyzer.},
  author={Ou, Xinming and Govindavajhala, Sudhakar and Appel, Andrew W and others},
  booktitle={USENIX security symposium},
  volume={8},
  pages={113--128},
  year={2005},
  organization={Baltimore, MD}
}

@article{pham2021survey,
  title={A survey on security attacks and defense techniques for connected and autonomous vehicles},
  author={Pham, Minh and Xiong, Kaiqi},
  journal={Computers \& Security},
  volume={109},
  pages={102269},
  year={2021},
  publisher={Elsevier}
}

@article{wang2024proactive,
  title={Proactive security defense: cyber threat intelligence modeling for connected autonomous vehicles},
  author={Wang, Yinghui and Ren, Yilong and Cui, Zhiyong and Yu, Haiyang},
  journal={arXiv preprint arXiv:2410.16016},
  year={2024}
}

@inproceedings{yang2023cakg,
  title={CAKG: A Framework for Cybersecurity Threat Detection of Automotive via Knowledge Graph},
  author={Yang, Peng and Lijie, Wang and Yun, Li and Xuedong, Song and Yaxin, Wang and Biheng, Guo},
  booktitle={2023 8th International Conference on Data Science in Cyberspace (DSC)},
  pages={221--228},
  year={2023},
  organization={IEEE}
}

@inproceedings{syed2016uco,
  title={UCO: A unified cybersecurity ontology},
  author={Syed, Zareen and Padia, Ankur and Mathews, M Lisa and Finin, Tim and Joshi, Anupam and others},
  booktitle={Proceedings of the AAAI Workshop on Artificial Intelligence for Cyber Security},
  pages={195--202},
  year={2016}
}

@article{injy2021open,
  title={Open-CyKG: An Open Cyber Threat Intelligence Knowledge Graph},
  author={Injy, Sarhan and Marco, Spruit},
  journal={Knowledge-Based Systems},
  volume={233},
  number={1},
  pages={1--13},
  year={2021}
}

@inproceedings{li2022attackg,
  title={AttacKG: Constructing technique knowledge graph from cyber threat intelligence reports},
  author={Li, Zhenyuan and Zeng, Jun and Chen, Yan and Liang, Zhenkai},
  booktitle={European Symposium on Research in Computer Security},
  pages={589--609},
  year={2022},
  organization={Springer}
}

@article{hu2024llm,
  title={Llm-tikg: Threat intelligence knowledge graph construction utilizing large language model},
  author={Hu, Yuelin and Zou, Futai and Han, Jiajia and Sun, Xin and Wang, Yilei},
  journal={Computers \& Security},
  volume={145},
  pages={103999},
  year={2024},
  publisher={Elsevier}
}

@inproceedings{cheng2025ctinexus,
  title={Ctinexus: Automatic cyber threat intelligence knowledge graph construction using large language models},
  author={Cheng, Yutong and Bajaber, Osama and Tsegai, Saimon Amanuel and Song, Dawn and Gao, Peng},
  booktitle={2025 IEEE 10th European Symposium on Security and Privacy (EuroS\&P)},
  pages={923--938},
  year={2025},
  organization={IEEE}
}

@article{chlup2022threatget,
  title={THREATGET: towards automated attack tree analysis for automotive cybersecurity},
  author={Chlup, Sebastian and Christl, Korbinian and Schmittner, Christoph and Shaaban, Abdelkader Magdy and Schauer, Stefan and Latzenhofer, Martin},
  journal={Information},
  volume={14},
  number={1},
  pages={14},
  year={2022},
  publisher={MDPI}
}

@inproceedings{huang2024ctikg,
  title={Ctikg: Llm-powered knowledge graph construction from cyber threat intelligence},
  author={Huang, Liangyi and Xiao, Xusheng},
  booktitle={First Conference on Language Modeling},
  year={2024}
}

@inproceedings{della2025cti,
  title={Cti-hal: A human-annotated dataset for cyber threat intelligence analysis},
  author={Della Penna, Sofia and Natella, Roberto and Orbinato, Vittorio and Parracino, Lorenzo and Pianese, Luciano},
  booktitle={2025 IEEE European Symposium on Security and Privacy Workshops (EuroS\&PW)},
  pages={69--78},
  year={2025},
  organization={IEEE}
}

@article{zhang2025attackg+,
  title={AttacKG+: Boosting attack graph construction with large language models},
  author={Zhang, Yongheng and Du, Tingwen and Ma, Yunshan and Wang, Xiang and Xie, Yi and Yang, Guozheng and Lu, Yuliang and Chang, Ee-Chien},
  journal={Computers \& Security},
  volume={150},
  pages={104220},
  year={2025},
  publisher={Elsevier}
}

@article{zhang2025mm,
  title={MM-AttacKG: A Multimodal Approach to Attack Graph Construction with Large Language Models},
  author={Zhang, Yongheng and Zhao, Xinyun and Ma, Yunshan and Ma, Haokai and Guan, Yingxiao and Yang, Guozheng and Lu, Yuliang and Wang, Xiang},
  journal={arXiv preprint arXiv:2506.16968},
  year={2025}
}

@article{wang2025dataset,
  title={A dataset for cyber threat intelligence modeling of connected autonomous vehicles},
  author={Wang, Yinghui and Ren, Yilong and Qin, Hongmao and Cui, Zhiyong and Zhao, Yanan and Yu, Haiyang},
  journal={Scientific Data},
  volume={12},
  number={1},
  pages={366},
  year={2025},
  publisher={Nature Publishing Group UK London}
}

@book{salfer2024automotive,
  title={Automotive Security Analyzer for Exploitability Risks},
  author={Salfer, Martin},
  year={2024},
  publisher={Springer}
}

@misc{autoisac_atm,
  title = {Auto-ISAC ATM},
  howpublished = {\url{https://atm.automotiveisac.com/}},
  author={Auto-ISAC},
  note = {Accessed: 25 Feb 2026}
}

@article{edge2024local,
  title={From local to global: A graph rag approach to query-focused summarization},
  author={Edge, Darren and Trinh, Ha and Cheng, Newman and Bradley, Joshua and Chao, Alex and Mody, Apurva and Truitt, Steven and Metropolitansky, Dasha and Ness, Robert Osazuwa and Larson, Jonathan},
  journal={arXiv preprint arXiv:2404.16130},
  year={2024}
}

@misc{openrouter,
  title = {OpenRouter API},
  howpublished = {\url{https://openrouter.ai/}},
  author={OpenRouter},
  note = {Accessed: 25 Feb 2026}
}

\appendix 
\section{Appendix}

\subsection{Ablation Study: Retrieval Modality}
\label{app:ablation}

Table~\ref{tab:ablation-overall} reports overall PF/KR scores across four retrieval conditions, and Table~\ref{tab:ablation-case} provides a per-case breakdown.

\begin{table}[h]
\centering
\caption{Ablation study: overall mean PF and KR across retrieval conditions (320 experiments each).}
\label{tab:ablation-overall}
\small
\begin{tabular}{lcc}
\toprule
\textbf{Condition} & \textbf{Mean PF} & \textbf{Mean KR} \\
\midrule
Hybrid (baseline)    & 46.6 & 54.5 \\
Vector-only (NO\_GRAPH) & 47.7 & 55.2 \\
Graph-only (NO\_VECTOR)  & 47.8 & 55.2 \\
NO\_RAG              & 47.8 & 54.9 \\
\bottomrule
\end{tabular}
\end{table}

\begin{table}[h]
\centering
\caption{Ablation study: per-case mean PF scores.}
\label{tab:ablation-case}
\small
\resizebox{\columnwidth}{!}{%
\begin{tabular}{lcccc}
\toprule
\textbf{Case} & \textbf{Hybrid} & \textbf{Vector-only} & \textbf{Graph-only} & \textbf{NO\_RAG} \\
\midrule
Jeep   & 33.9 & 40.1 & 33.2 & 38.6 \\
MBUX   & 47.1 & 46.2 & 46.7 & 41.6 \\
Nissan & 39.3 & 40.0 & 39.8 & 40.6 \\
BMW    & 66.4 & 64.4 & 71.5 & 70.6 \\
\bottomrule
\end{tabular}%
}
\end{table}

No condition consistently outperforms the others; per-case effects are heterogeneous (\textit{e.g.}, MBUX drops under NO\_RAG while BMW improves). Notably, the Hybrid baseline records a marginally lower overall PF (46.6) than single-modality or No-RAG conditions (47.7--47.8); however, all differences fall within $\pm$1.2 points---well below the 95\% CI of $\pm$5.9 reported in RQ2---indicating that these variations are not statistically meaningful. These results support the Discussion finding that knowledge granularity, not retrieval modality, is the dominant performance factor.

\subsection{Prompt Templates}
\label{app:prompts}

This section presents the core prompt templates used in the KB construction and attack graph generation stages.

\subsubsection{Entity Extraction Prompt}
The entity extraction prompt assigns the LLM the role of an automotive cybersecurity analyst and applies three constraints: (1)~\textbf{Verbatim Extraction}---all entities must be explicitly grounded in the input text; (2)~\textbf{Negative Instruction}---entities not mentioned in the text must not be created; and (3)~\textbf{Structural Guidance}---12 entity types are defined with explicit examples to reduce boundary ambiguity. Key excerpts:

\begin{mdframed}[backgroundcolor=black!5, linewidth=0.5pt, innertopmargin=4pt, innerbottommargin=4pt, innerleftmargin=4pt, innerrightmargin=4pt]
\begin{lstlisting}
You are an expert automotive cybersecurity analyst
specializing in extracting structured information
from technical texts, incident reports, and
vulnerability descriptions.

### Additional Instructions for Accuracy:
- Grounding: Ensure all identified entities are
  derived explicitly from the input text.
  Do not hallucinate entities not present in the text.
- Standardization: If an entity is referred to
  multiple times with slight variations, always
  standardize to the exact same name.
- Coreference Resolution: If a pronoun is used,
  replace it with the specific entity name it
  refers to. Do NOT extract pronouns as entities.
- Conciseness: Entities must be specific noun
  phrases (typically 1-5 words). Do NOT extract
  full sentences or descriptions of actions.

### Negative Instructions:
- Do NOT extract or infer relationships between
  entities.
- Do NOT create entities that are not explicitly
  mentioned in the text.
- Do NOT extract generic terms like "device",
  "system", "vehicle" unless part of a specific
  name (\textit{e.g.}, "Infotainment system").
- Do NOT extract verbs or action phrases
  (\textit{e.g.}, "hijack the connection") as entities.

### Entity Definitions (with examples):
- component: (\textit{e.g.}, 'UDS module', 'BT stack')
- target_ecu: (\textit{e.g.}, 'Infotainment ECU')
- vulnerability: (\textit{e.g.}, 'Buffer overflow (CWE-120)')
                ...[snip 9 more types]...

### Output Format:
{ "component": [], "target_ecu": [],
  "vulnerability": [], ... }
\end{lstlisting}
\end{mdframed}

\subsubsection{Relation Extraction Prompt}
The relation extraction prompt constrains outputs to nine predefined relation types (Table~\ref{tab:relation_types}) and requires that both source and target entities come verbatim from the previously extracted entity list ($E_{extracted}$):

\begin{mdframed}[backgroundcolor=black!5, linewidth=0.5pt, innertopmargin=4pt, innerbottommargin=4pt, innerleftmargin=4pt, innerrightmargin=4pt]
\begin{lstlisting}
Extract valid relationships only between the
entities in "Extracted Entities".
Always use entity strings verbatim from that list
for each "source" and "target".

### Allowed Relations
(Each line: Description - Format - Example)

1. HAS_VULNERABILITY
2. IDENTIFIED_BY
3. CONNECTED_TO
4. USES_PROTOCOL
5. DELIVERED_VIA
6. PROVIDED_BY
7. REPORTED_BY
8. AFFECTS_ASSET
9. USED_AGAINST
                ...[snip per-relation examples]...

If no valid relations exist, return "Relations": [].
\end{lstlisting}
\end{mdframed}

\subsubsection{Rewriting Prompt}
The rewriting prompt converts structured data (entities and relations) into dense, factual natural language paragraphs optimized for vector retrieval. It aims to align semantically with ATM technique descriptions by emphasizing attack mechanisms and impacts:

\begin{mdframed}[backgroundcolor=black!5, linewidth=0.5pt, innertopmargin=4pt, innerbottommargin=4pt, innerleftmargin=4pt, innerrightmargin=4pt]
\begin{lstlisting}
You are an expert in automotive cybersecurity
documentation. Your task is to rewrite the provided
structured data (Entities and Relations) into a
dense, factual paragraph optimized for vector
retrieval.

### Goal
Create a text that semantically aligns with
Automotive Threat Matrix (ATM) technique
descriptions. Describe the mechanism of the attack
(how it works) and the impact (what it achieves).

### Rules
1. Action-Oriented: Use strong verbs describing
   the attack lifecycle (\textit{e.g.}, "intercepted",
   "spoofed", "flooded", "extracted", "modified").
2. Mechanism & Impact: Explicitly state HOW the
   vulnerability was exploited and WHAT the
   consequence was.
3. Keyword Preservation: Keep specific entity
   names (ECUs, Protocols, CVEs) exact.
4. Conciseness: Dense, technical summary.
5. Contextual Linking: Connect the 'Tool' used
   to the 'Vulnerability' exploited and the
   'Asset' affected.
6. Grounding: Do not invent details not present
   in the input.
                ...[snip example and I/O format]...
\end{lstlisting}
\end{mdframed}

\subsubsection{SFT Data Generation Prompt}
The SFT data generation prompt assigns the LLM the role of a Threat Intelligence Analyst. It instructs the model to synthesize virtual incident reports for each ATM technique across five styles with diverse scenarios. To prevent label leakage, technique names and IDs are explicitly blacklisted from the generated text:

\begin{mdframed}[backgroundcolor=black!5, linewidth=0.5pt, innertopmargin=4pt, innerbottommargin=4pt, innerleftmargin=4pt, innerrightmargin=4pt]
\begin{lstlisting}
You are an expert Threat Intelligence Analyst
specializing in automotive cybersecurity.
Your mission is to draft five distinct reports
that illustrate the SAME Auto-ISAC attack
technique through different scenarios.

## Safety & Ethics (MANDATORY)
- Keep tone professional and avoid sensationalism.

## Global Requirements
- All five reports must express the same technique,
  each via a distinct scenario, attacker type,
  target surface, and network domain.
- Each report must be 3-6 paragraphs.
- Focus on impacts, detection, and mitigations
  over exploitation details.
- IMPORTANT: The narrative text MUST NOT contain
  or repeat any of the following terms (to
  prevent label leakage): [blocked_terms]

## Output Schema
- "style": one of ["research_lab",
  "corporate_advisory", "press_release",
  "anon_forum_post", "training_material"]
                ...[snip 10 more fields]...

## Diversity Constraints
- Vary at least three of: attacker_type,
  target_surface, network_domain,
  kill-chain entry points.
\end{lstlisting}
\end{mdframed}

\subsubsection{Graph Summarization Prompt}
For densely-connected nodes, the following prompt condenses raw SPO triples into a security summary:

\begin{mdframed}[backgroundcolor=black!5, linewidth=0.5pt, innertopmargin=4pt, innerbottommargin=4pt, innerleftmargin=4pt, innerrightmargin=4pt]
\begin{lstlisting}
You are an expert Automotive Cybersecurity Analyst
and Knowledge Graph Summarizer.

Analyze the provided SPO triplets connected to
a single Hub Node. Generate a summary paragraph
(max 80 words) capturing the core threat landscape.
                ...[snip detailed instructions]...

Focus on:
- Attack Categories: Group vulnerabilities into
  2-3 high-level categories (\textit{e.g.}, Authentication
  Bypass, Memory Corruption, Protocol Flaws).
- Impact: Primary exploit capabilities
  (\textit{e.g.}, RCE, Physical Access).
- Context: Maintain automotive domain terminology.

HUB NODE: {entity_name}
--- RAW TRIPLET DATA ---
{relations}
\end{lstlisting}
\end{mdframed}

\subsubsection{Attack Graph Generation Prompt}
The final inference prompt integrates both retrieval sources (VectorDB and GraphDB) into a unified context, with explicit conflict resolution rules and a structured JSON output schema:

\begin{mdframed}[backgroundcolor=black!5, linewidth=0.5pt, innertopmargin=4pt, innerbottommargin=4pt, innerleftmargin=4pt, innerrightmargin=4pt]
\begin{lstlisting}
You are an expert automotive cybersecurity analyst.

=== VEHICLE TOPOLOGY ===
[ECU Details, External Interfaces,
 Network Connections from vehicle spec]

=== RETRIEVED KNOWLEDGE ===
Similar Attack Cases: [VectorDB results]
Graph Context: [GraphDB traversal results]
                ...[snip]...

=== CONFLICT RESOLUTION RULE ===
If VectorDB (\textit{e.g.}, patch notes) conflicts with
GraphDB (\textit{e.g.}, existing vulnerability):
1. Prioritize VectorDB ONLY IF it explicitly
   mentions the target vehicle's make, model,
   and year.
2. Otherwise, assume the vulnerability exists
   (Conservative Approach).

=== OUTPUT SCHEMA ===
{
  "nodes": [{"id", "label", "type", "group"}],
  "edges": [{"source", "target", "step_id",
             "type", "label", "protocol",
             "technique_ids": ["ATM-Txxxx"]}]
}
                ...[snip requirements]...
\end{lstlisting}
\end{mdframed}

\subsection{Representative Attack Graphs}

\begin{landscape}
\begin{figure}[p]
    \centering
    \includegraphics[width=\linewidth]{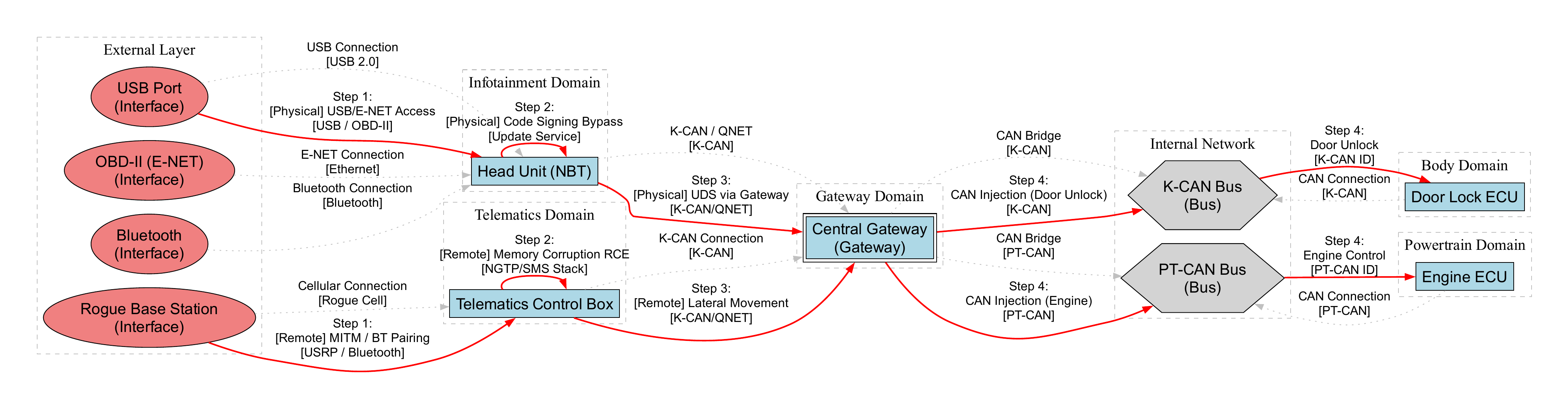}
    \vspace{1em}
    \includegraphics[width=\linewidth]{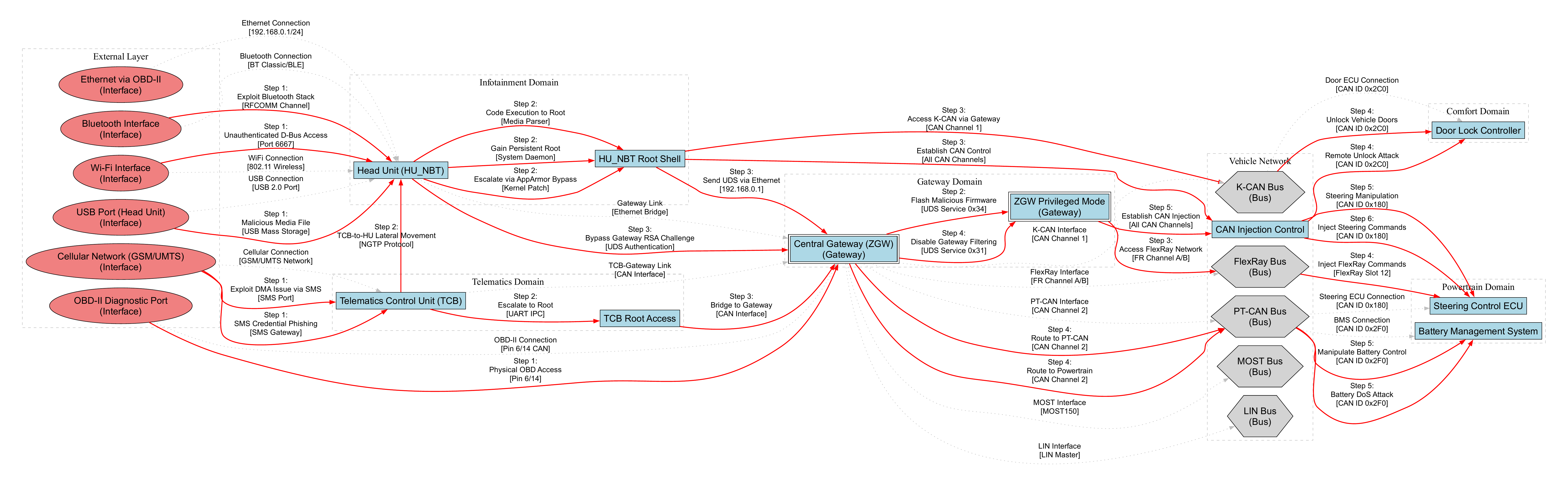}
    \caption{Comparison of attack graphs for Case 4 (BMW i3). Top:~Ground-truth---two validated paths (Physical via USB/OBD-II and Remote via rogue base station) converging at the Central Gateway. Bottom:~LLM-generated (Claude 4.5 Sonnet, LOO condition)---the model reconstructs both GT paths while exploring additional entry surfaces and alternative attack techniques.}
    \label{fig:ag_comparison_bmw}
\end{figure}
\end{landscape}

\clearpage

\begin{landscape}
\begin{figure}[p]
    \centering
    \includegraphics[width=\linewidth]{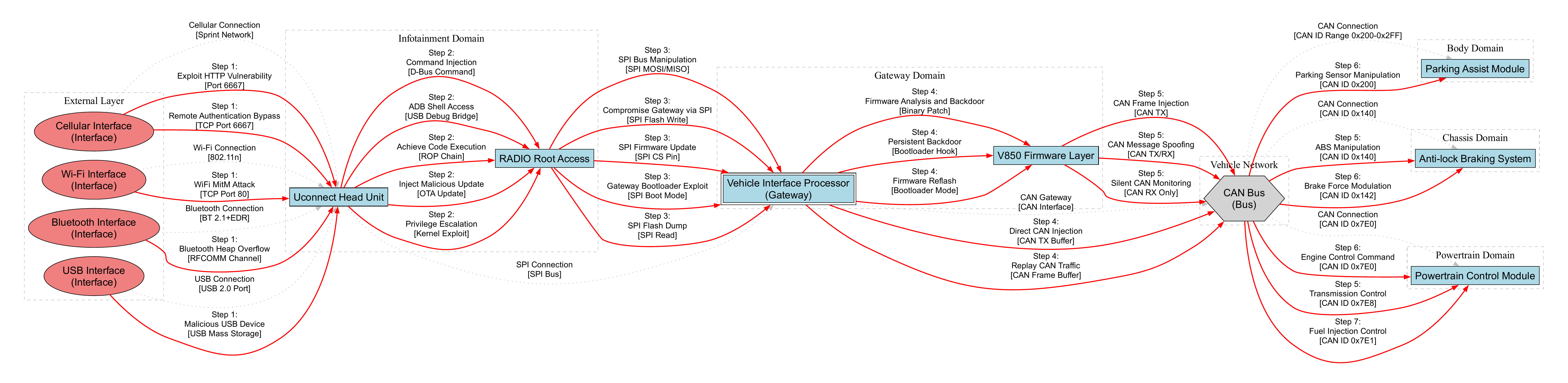}
    \caption{Best-scoring LLM-generated attack graph for Case~1 (Jeep Cherokee). Generated by Claude~4.5~Sonnet, Iteration~4 (PF\,=\,78.0, KR\,=\,73.8). The graph comprises 12~nodes and 39~edges, identifying 5~distinct attack paths from 4~entry points (Cellular, WiFi, Bluetooth, USB) converging through the RADIO head unit and V850 gateway to reach safety-critical targets (PAM, ABS, PCM) via the CAN bus. Red edges denote attack steps; gray edges denote network connections.}
    \label{fig:best_ag_jeep}
\end{figure}
\end{landscape}

\begin{landscape}
\begin{figure}[p]
    \centering
    \includegraphics[width=\linewidth]{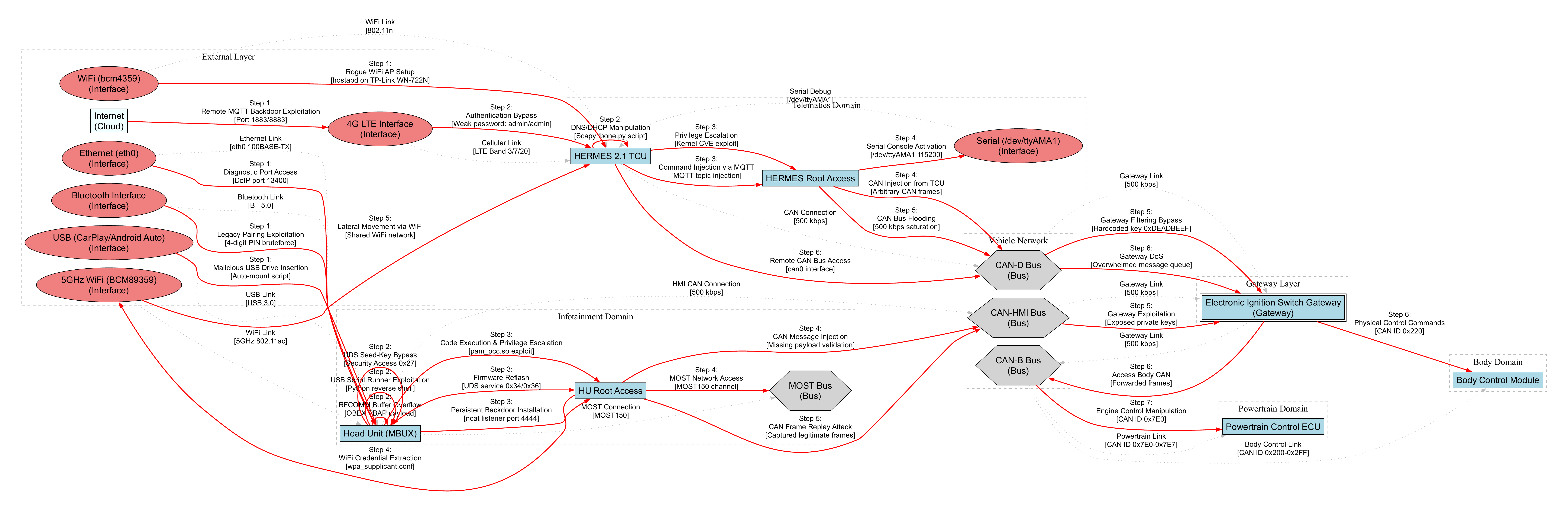}
    \caption{Best-scoring LLM-generated attack graph for Case~2 (Mercedes MBUX). Generated by Claude~4.5~Sonnet, Iteration~8 (PF\,=\,83.5, KR\,=\,80.7). The graph comprises 20~nodes and 52~edges, identifying 4~major attack paths from 8~entry points (4G\_LTE, WiFi, Bluetooth, USB, Ethernet, Serial, Internet) through the T-Box (HERMES) and HU (NTG6) to reach Powertrain\_ECU and Body\_Control\_ECU via multi-bus pivoting across CAN\_D, CAN\_HMI, and CAN\_B.}
    \label{fig:best_ag_mbux}
\end{figure}
\end{landscape}

\begin{landscape}
\begin{figure}[p]
    \centering
    \includegraphics[width=\linewidth]{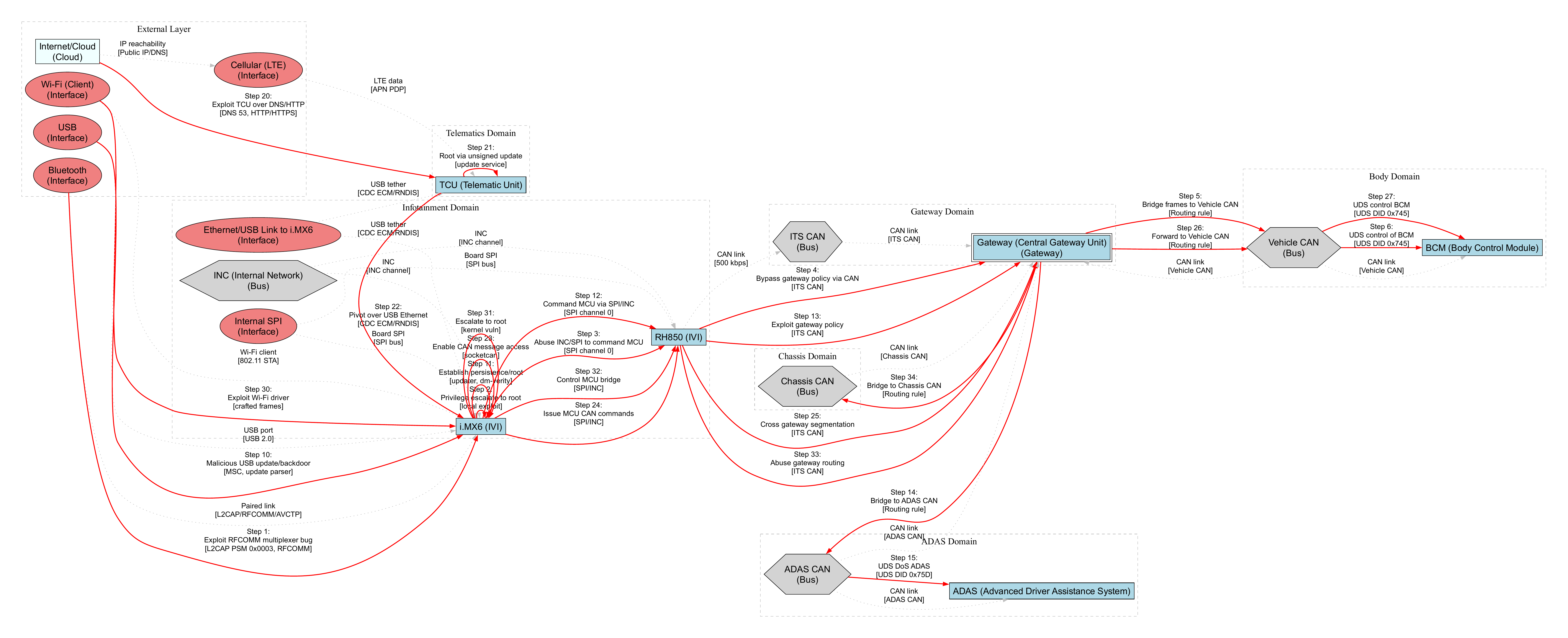}
    \caption{Best-scoring LLM-generated attack graph for Case~3 (Nissan Leaf). Generated by GPT~5.2, Iteration~10 (PF\,=\,78.0, KR\,=\,73.0). The graph comprises 18~nodes and 47~edges, identifying 4~primary attack paths from 5~entry points (Bluetooth, WiFi, USB, Cellular\_LTE, Internet). The longest chain spans 9~steps from Internet through TCU and iMX6\_IVI to BCM. The graph highlights ADAS-specific targeting via dedicated ADAS\_CAN and weak gateway segmentation across 5~CAN buses.}
    \label{fig:best_ag_nissan}
\end{figure}
\end{landscape}

\begin{landscape}
\begin{figure}[p]
    \centering
    \includegraphics[width=\linewidth]{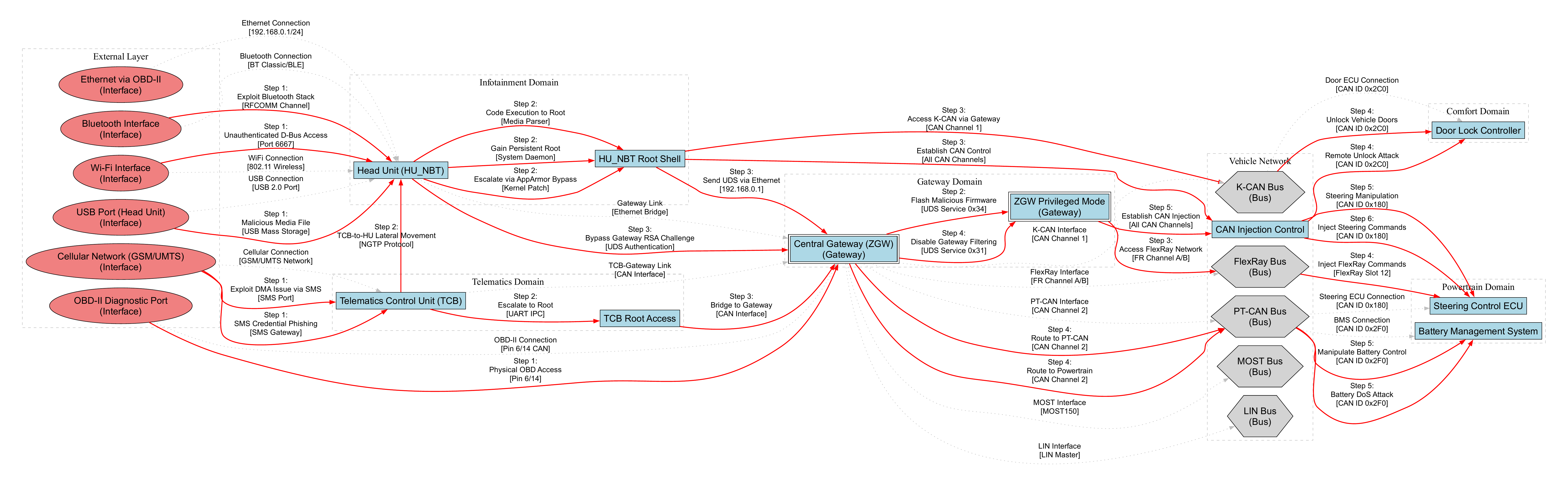}
    \caption{Best-scoring LLM-generated attack graph for Case~4 (BMW i3). Generated by Claude~4.5~Sonnet, Iteration~3 (PF\,=\,88.5, KR\,=\,87.5). The most complex graph among the four cases, comprising 21~nodes and 59~edges with 5~distinct attack paths from 6~entry points (Cellular, WiFi, Bluetooth, USB, OBD-II, Ethernet). The graph spans a multi-bus architecture (K-CAN, PT-CAN, FlexRay, MOST, LIN) converging through the ZGW central gateway to reach Steering\_ECU, BMS\_ECU, and Door\_Lock\_ECU.}
    \label{fig:best_ag_bmw}
\end{figure}
\end{landscape}

\end{document}